\documentclass[onecolumn, draftcls, 11pt]{IEEEtran}
\usepackage{amsmath, amssymb, amsfonts}
\usepackage{graphicx}
\usepackage{epsfig}
\usepackage{caption}
\usepackage{cases}
\usepackage{tensor}
\usepackage{fancyhdr}
\usepackage{setspace}

\usepackage{color}
\usepackage{subcaption}

\makeatletter
\newcommand{\vast}{\bBigg@{4}}
\newcommand{\Vast}{\bBigg@{5}}
\makeatother

\title{Effect of Receive Spatial Diversity on the Degrees of Freedom Region in Multi-Cell Random Beamforming
\footnote{This work was presented in part at {\it IEEE Wireless Communications and Networking Conference (WCNC)}, Shanghai, China, April 07-10, 2013.} 
\footnote{The authors are with the Department of Electrical and Computer Engineering, National University of Singapore (email(s): \{hieudn, elezhang, elehht\}@nus.edu.sg).}}
\author{Hieu Duy Nguyen, Rui Zhang, and Hon Tat Hui}

\begin{document}
\maketitle \thispagestyle{empty}

\begin{abstract}
The \emph{random beamforming} (RBF) scheme, jointly applied with \emph{multi-user diversity} based scheduling, is able to achieve virtually interference-free downlink transmissions with only \emph{partial} channel state information (CSI) available at the transmitter. However, the impact of receive spatial diversity on the rate performance of RBF is not fully characterized yet even in a single-cell setup. In this paper, we study a multi-cell multiple-input multiple-output (MIMO) broadcast system with RBF applied at each base station (BS) and either the minimum-mean-square-error (MMSE), matched filter (MF), or antenna selection (AS) based spatial receiver employed at each mobile terminal. We investigate the effect of different spatial diversity receivers on the achievable sum-rate of multi-cell RBF systems subject to both the intra- and inter-cell interferences. We first derive closed-form expressions for the distributions of the receiver signal-to-interference-plus-noise ratio (SINR) with different spatial diversity techniques, based on which we compare their rate performances at finite signal-to-noise ratios (SNRs). We then investigate the asymptotically high-SNR regime and for a tractable analysis assume that the number of users in each cell scales in a certain order with the per-cell SNR as SNR goes to infinity. Under this setup, we characterize the degrees of freedom (DoF) region for multi-cell RBF systems with different types of spatial receivers, which consists of all the achievable DoF tuples for the individual sum-rate of all the cells. The DoF region analysis provides a succinct characterization of the interplays among the receive spatial diversity, multiuser diversity, spatial multiplexing gain, inter-/intra-cell interferences, and BSs' collaborative transmission. 
\end{abstract}




\setlength{\baselineskip}{1.1\baselineskip}
\markboth{IEEE Transactions on Wireless Communications}%
{Submitted paper}

\newtheorem{definition}{\underline{Definition}}[section]
\newtheorem{fact}{Fact}
\newtheorem{assumption}{\underline{Assumption}}[section]
\newtheorem{theorem}{\underline{Theorem}}[section]
\newtheorem{lemma}{\underline{Lemma}}[section]
\newtheorem{corollary}{\underline{Corollary}}[section]
\newtheorem{proposition}{\underline{Proposition}}[section]
\newtheorem{example}{\underline{Example}}[section]
\newtheorem{remark}{\underline{Remark}}[section]
\newtheorem{algorithm}{\underline{Algorithm}}[section]
\newcommand{\mv}[1]{\mbox{\boldmath{$ #1 $}}}

\newpage
\section{Introduction}\label{sec:introduction}

Recent advance in wireless communication has shifted from single-user multiple-input multiple-output (MIMO) to multi-user MIMO systems, which greatly enhances the performance by transmitting to multiple users simultaneously via spatial multiplexing. The capacity region of a single-cell MU-MIMO downlink system, also called MIMO broadcast channel (BC), is achieved by the non-linear ``Dirty Paper Coding (DPC)'' scheme \cite{Weingarten01}, which is of high implementation complexity. Other studies have thus proposed low-complexity linear MIMO precoder designs, e.g., block diagonalization \cite{Spencer01}. A common drawback of the above schemes is the requirement of instantaneous and highly accurate channel state information (CSI) at the transmitter, which is practically difficult to realize.

The single-beam ``opportunistic beamforming (OBF)'' and multi-beam ``random beamforming (RBF)'' schemes for the single-cell multiple-input single-output (MISO) BC, introduced in \cite{Viswanath01} and \cite{Sharif01}, respectively, have attracted a great deal of attention since they require only \emph{partial} CSI fedback to the transmitter at each base station (BS). The fundamental idea in these schemes is to achieve nearly interference-free downlink transmissions by exploiting the multi-user channel diversity with \emph{opportunistic} user scheduling. It has been shown that the achievable sum-rates with the RBF and optimal DPC both scale identically as the number of users in the cell approaches infinity, for any given signal-to-noise ratio (SNR) \cite{Sharif01}, \cite{Sharif02}. This shows the optimality of RBF in the regime of large number of users and has motivated extensive subsequent studies on, e.g., sum-rate characterization \cite{Kim01}, \cite{Park02}, quantized channel feedback \cite{Ozdemir01}-\cite{Sanayei01}, and precoder design with opportunistic scheduling \cite{Louie01}, \cite{Yoo01}. However, most existing studies of RBF have only considered the single-cell setup. One recent progress was made in our prior work \cite{Hieu01}, in which the rate performance of multi-cell MISO RBF systems is investigated in both the finite-SNR and asymptotically high-SNR regimes.

Furthermore, the effect of \emph{receive spatial diversity} on the rate performance of RBF with multi-antenna receivers is not yet fully characterized in the literature, even in the single-cell case. Note that some prior works have studied RBF under a single-cell MIMO setup, e.g., \cite{Sharif01}, \cite{Sharif02}. Assuming that the number of users goes to infinity for any given SNR, it has been shown therein that RBF schemes with single- or multi-antenna receivers achieve the same sum-rate scaling law with the growing number of users. The conventional asymptotic analysis thus leads to a pessimistic result that receive spatial diversity provides only \emph{marginal} gains to the achievable rate of RBF \cite{Sharif01}, \cite{Sharif02}. In contrast, in this paper, we investigate the achievable rate of a multi-cell MIMO RBF system with different receive spatial diversity techniques in the \emph{high-SNR} regime. We aim to characterize the achievable degrees of freedom (DoF) trade-offs in multi-cell MIMO RBF systems, where the DoF is defined as the individual sum-rate of each cell normalized by the logarithm of the per-cell SNR as SNR goes to infinity. Thereby, we provide a succinct characterization of the interplays among the receive spatial diversity, multiuser diversity, spatial multiplexing gain, inter-/intra-cell interferences, and BSs' collaborative transmission in multi-cell RBF systems.

It is worth noting that the high-SNR DoF analysis for interference channels has become a major research topic inspired by the invention of a novel transmission technique so-called ``interference alignment (IA)'' (see, e.g., \cite{Jafar01} and references therein). Although IA-based DoF studies provide useful insights to the optimal transmission design for interference-limited multi-cell systems, they have in general assumed the perfect CSI at BSs. Furthermore, how to efficiently schedule the users' transmissions in IA-based systems with a significantly larger number of users than that of transmitting antennas remains open. Some promising results on this regard can be found in \cite{Jung01}-\cite{Lee01} and the references therein. Investigation on the multi-cell cooperative downlink precoding/beamforming at finite SNRs has also been pursued in the literature under two different assumptions on the cooperation level among BSs, i.e., the ``fully cooperative'' multi-cell systems with global transmit messages sharing across all BSs \cite{Ng01}-\cite{ZhangLan} and ``partially cooperative ''counterparts with only locally available transmit message at each BS \cite{Dahrouj01}-\cite{Liu01}. Furthermore, there has been recent work on the asymptotic analysis for multi-cell MIMO downlink systems based on a large-system approach, in which the number of users per cell and the number of transmit antennas per BS both go to infinity at the same time with a fixed ratio \cite{Huh01}, \cite{Zakhour01}. 

The main results of this paper are summarized as follows.
\begin{itemize}
	\item {\bf Multi-cell MIMO RBF}: We propose three MIMO RBF schemes for multi-cell downlink systems. In these schemes, RBF is applied at each BS and either the minimum-mean-square-error (MMSE), matched filter (MF), or antenna selection (AS) based spatial receiver is employed at each mobile terminal (denoted as RBF-MMSE, RBF-MF, and RBF-AS schemes, respectively). These schemes preserve the same low-feedback requirement as that for the special case of single-cell OBF/RBF \cite{Viswanath01}, \cite{Sharif01}, but bring in the new benefits of receive spatial diversity with different performance-complexity trade-offs. 
	
	\item {\bf SINR Distribution}: By applying the tools from \emph{multivariate analysis} (MVA), we derive the exact distribution of the signal-to-interference-plus-noise ratio (SINR) at each multi-antenna receiver in a multi-cell MIMO RBF system subject to both the intra- and inter-cell interferences, assuming either the MMSE, MF, or AS based spatial diversity technique. Note that these results are non-trivial extensions of our previous work \cite{Hieu01} for the MISO RBF case with only single-antenna receivers.	
	
	\item {\bf DoF Region Characterization}: We further investigate the multi-cell MIMO RBF system with MMSE, MF, or AS based spatial receivers under the asymptotically high-SNR regime, by assuming that the number of users per cell scales in a certain order with the SNR (a larger scaling order indicates a higher user density in one particular cell). We first derive the achievable sum-rate DoF under a single-cell setup without the inter-cell interference to gain useful insights and then obtain a general characterization of the DoF region for the multi-cell case, which constitutes all the achievable DoF tuples for the individual sum-rate of all the cells subject to the additional inter-cell interference. Our analysis reveals that significant sum-rate DoF gains can be achieved by employing the MMSE-based spatial receiver as compared to the cases with single-antenna receivers or with the suboptimal spatial receivers such as MF and AS. This is in sharp contrast to the existing result (e.g., \cite{Sharif01}, \cite{Sharif02}) that spatial diversity receivers only yield marginal rate gains in RBF, which is based on the conventional asymptotic analysis in the regime of large number of users but with fixed SNR per cell. With MMSE receivers, our result shows that a significantly less number of users in each cell is required to achieve a given sum-rate DoF target as compared to the cases without receiver spatial diversity or with MF/AS receivers. Our new high-SNR DoF analysis thus provides a more realistic characterization of the rate trade-offs in multi-cell MIMO RBF systems.
\end{itemize}

The remainder of this paper is organized as follows. Section \ref{sec:system model} describes the multi-cell MIMO downlink system model and the MIMO RBF scheme with MMSE, MF, or AS based spatial receivers. Section \ref{sec:SINR distributions} investigates the SINR distribution in each receiver case based on MVA. Section \ref{sec:DoF characterization} characterizes the achievable sum-rate DoF for single-cell MIMO RBF, and then extends the result to the DoF region characterization for multi-cell MIMO RBF. Finally, we conclude the paper in Section \ref{sec:conclusion}. 

{\it Notations}: Scalars, vectors, and matrices are denoted by lower-case, bold-face lower-case, and bold-face upper-case letters, respectively. The matrix transpose and conjugate transpose operators are denoted as $(\cdot)^T$ and $(\cdot)^H$, respectively. $\mathbb{E}[\cdot]$ denotes the statistical expectation. $\mv{Tr}(\cdot)$ represents the trace of a matrix. The distribution of a circularly symmetric complex Gaussian (CSCG) random vector with mean vector $\mv{a}$ and covariance matrix $\mv{\Sigma}$ is denoted by $\mathcal{CN}(\mv{a},\mv{\Sigma})$. The notation $\sim$ stands for ``distributed as''. $\mathbb{C}^{x \times y}$ denotes the space of $x\times y$ complex matrices. $\mv{I}_p$ and $\mv{0}_{p\times n}$ stand for the identity matrix and the all-zero matrix with the corresponding dimensions, respectively. We use $|a|$ to represent the magnitude of a complex number $a$. A diagonal matrix with diagonal elements $a_1, \dots, a_n$ is denoted as $diag(a_1,\ldots,a_n)$.

\section{System Model}\label{sec:system model}

Consider a cellular system consisting of $C$ cells and $K_c$ mobile stations (MSs) in the $c$-th cell, $c=1,\cdots,C$. In this paper, we focus on the downlink transmission assuming universal frequency reuse, i.e., all cells are assigned the same bandwidth for transmission. For the ease of analysis, we also assume that all BSs/MSs have the same number of transmit/receive antennas, denoted as $N_T$ and $N_R$, respectively. Consider time-slotted transmissions, at each time slot, the $c$-th BS transmits $M_c$ orthonormal beams and selects $M_c$ users in the $c$-th cell for transmission, with $M_c\leq N_T$ and $M_c\leq K_c$, $\forall c$. The received baseband signal of user $k$ in the $c$-th cell is given by
	\begin{align}\label{eq:signal model}
	\mv{y}_{k}^{(c)} = \mv{H}_{k}^{(c,c)}\sum_{m=1}^{M_c} \mv{\phi}_{m}^{(c)} {\it s}_m^{(c)} +\sum_{l=1,~l\neq c}^{C} \sqrt{\gamma_{l,c}} \mv{H}_{k}^{(l,c)}\sum_{m=1}^{M_l} \mv{\phi}_{m}^{(l)} {\it s}_m^{(l)} + \mv{z}_{k}^{(c)},
	\end{align}
where $\mv{H}_{k}^{(l,c)} \in \mathbb{C}^{N_R\times M_l}$ denotes the MIMO channel matrix from the $l$-th BS to the $k$-th user of the $c$-th cell, which is assumed to be independent and identically distributed (i.i.d.) Rayleigh fading, i.e., all elements are i.i.d. and have the same distribution $\mathcal{CN}(0,1)$; $\mv{\phi}_{m}^{(c)} \in \mathbb{C}^{M_c\times 1}$ and $\it{s}_m^{(c)}$ are the $m$-th randomly generated beamforming vector of unit norm and the corresponding transmitted data symbol from the $c$-th BS, respectively; it is assumed that each BS has an average sum power constraint, $P_T$, i.e., $\mv{Tr} \left(\mathbb{E}[\mv{s}_{c}\mv{s}_{c}^{H}] \right)$ $\leq$ $P_T$, where $\mv{s}_{c} = [\it{s}_1^{(c)},\cdots,\it{s}_{M_c}^{(c)}]^T$; $\gamma_{l,c}$ $< 1$ stands for the (more severe) signal attenuation from the $l$-th BS to any user in the $c$-th cell, $l\neq c$; and $\mv{z}_{k}^{(c)}$ $\in \mathbb{C}^{N_R\times 1}$ is the receiver additive white Gaussian noise (AWGN) vector, which consists of i.i.d. random variables each distributed as $\mathcal{CN}(0,\sigma^2)$, $\forall k,c$. In the $c$-th cell, the total SNR, the SNR per beam, and the interference-to-noise ratio (INR) per beam from the $l$-th cell, $l\neq c$, are denoted as $\rho=P_T/\sigma^2$, $\eta_c=P_T/(M_c\sigma^2)$, and $\mu_{l,c}=\gamma_{l,c}P_T/(M_l\sigma^2)$, respectively.

\subsection{Multi-Cell RBF}\label{subsec:RBF-MMSE and RBF-MF}
 						 					
With multiple receive antennas, each MS can apply spatial diversity techniques to enhance the performance. In this paper, we propose the optimal MMSE -based spatial receiver and two suboptimal spatial receivers based on MF and AS, respectively. We describe the multi-cell RBF scheme with MMSE, MF, and AS receivers as follows.
\begin{enumerate}
	\item Training phase: 
		\begin{enumerate}
			\item[a)] The $c$-th BS generates $M_c$ orthonormal beams, $\mv{\phi}_{1}^{(c)}$, $\cdots$,$\mv{\phi}_{M_c}^{(c)}$, and uses them to broadcast the training signals to all users in the $c$-th cell. The total power of the $c$-th BS is assumed to be distributed equally over $M_c$ beams.
			\item[b1)] RBF-MMSE: For each of the $M_c$ beams, user $k$ in the $c$-th cell does the following:
				\begin{enumerate}
					\item Estimate the effective channel with training from the $c$-th BS: $\mv{\tilde{h}}_{k,m}^{(c,c)}=\mv{H}_k^{(c,c)}\mv{\phi}_m^{(c)}$, $1\leq m\leq M_c$.
					\item Estimate the interference-plus-noise covariance matrix due to the other $M_c-1$ beams from the $c$-th BS and all beams from the other $C-1$ BSs: 
						\begin{align}\label{eq:INCM}
							\mv{W}_k^{(c)} = \frac{P_T}{M_c} \mv{\tilde{H}}_{k,-m}^{(c,c)} \Big(\mv{\tilde{H}}_{k,-m}^{(c,c)}\Big)^H + \displaystyle\sum_{l=1,l\neq c}^{C}\frac{P_T\gamma_{l,c}}{M_l}\mv{\tilde{H}}_{k}^{(l,c)}\Big(\mv{\tilde{H}}_{k}^{(l,c)}\Big)^H +\sigma^2\mv{I},
						\end{align}
					where $\mv{\tilde{H}}_{k,-m}^{(c,c)}$ = $\mv{H}_{k}^{(c,c)}$ [$\mv{\phi}_{1}^{(c)}$, $\cdots$, $\mv{\phi}_{m-1}^{(c)}$, $\mv{\phi}_{m+1}^{(c)}$, $\cdots$, $\mv{\phi}_{M_c}^{(c)}$], and $\mv{\tilde{H}}_{k}^{(l,c)}$ = $\mv{H}_{k}^{(c,c)}$ [$\mv{\phi}_{1}^{(l)}$, $\cdots$, $\mv{\phi}_{M_l}^{(l)}$].
					\item Apply the MMSE spatial receiver, i.e., $\mv{t}_{k,m}^{(c)}$ = $\sqrt{\frac{P_T}{M_c}}$ $\left( \mv{W}_k^{(c)} \right)^{-1}$ $\mv{\tilde{h}}_{k,m}^{(c,c)}$, $1\leq m\leq M_c$, and compute the SINR corresponding to the $m$-th beam $\mv{\phi}_m^{(c)}$, i.e.,
						\begin{align}\label{eq:SINR RBF-MMSE}
							\text{SINR}_{k,m}^{(\text{MMSE},c)} = \frac{P_T}{M_c}\Big(\mv{\tilde{h}}_{k,m}^{(c,c)}\Big)^H \left( \mv{W}_k^{(c)} \right)^{-1} 	\mv{\tilde{h}}_{k,m}^{(c,c)}.
						\end{align}
					\item Send back $\text{SINR}_{k,m}^{(\text{MMSE},c)}$, $1\leq m\leq M_c$, to the $c$-th BS.
				\end{enumerate}
			\item[b2)] RBF-MF: For each of the $M_c$ beams, user $k$ in the $c$-th cell does the following:
				\begin{enumerate}
					\item Estimate the effective channel with training from the $c$-th BS: $\mv{\tilde{h}}_{k,m}^{(c,c)} = \mv{H}_k^{(c,c)}\mv{\phi}_m^{(c)}$, $1\leq m\leq M_c$. 
					\item Apply the MF spatial receiver, i.e., $\mv{t}_{k,m}^{(c)}$ = $\mv{\tilde{h}}_{k,m}^{(c,c)}$ $/$ $||\mv{\tilde{h}}_{k,m}^{(c,c)}||$. The rationale is to maximize the power received from the $m$-th beam. The receiver output is given by
					\begin{align}\label{eq:MF effective channel model}
 						r_{k,m}^{(c)} & = \sqrt{\frac{P_T}{M_c}} \left(\mv{t}_{k,m}^{(c)}\right)^H \mv{\tilde{h}}_{k,m}^{(c,c)} {\it s}_m^{(c)} + \sqrt{\frac{P_T}{M_c}} \left(\mv{t}_{k,m}^{(c)}\right)^H \mv{\tilde{H}}_{k,-m}^{(c,c)} \mv{s}_{-m}^{(c)} + \notag \\ 
						& \qquad \qquad \qquad \sum_{l=1, l\neq c} \sqrt{\frac{P_T\gamma_{l,c}}{M_l}} \left(\mv{t}_{k,m}^{(c)}\right)^H \mv{\tilde{H}}_{k}^{(l,c)} \mv{s}_{l} + \left(\mv{t}_{k,m}^{(c)}\right)^H \mv{z}_{k}^{(c)},
 						\end{align}
					where $\mv{s}_{-m}^{(c)}$ = [$s_{1}^{(c)}$, $\cdots$, $s_{m-1}^{(c)}$, $s_{m+1}^{(c)}$, $\cdots$, $s_{M_c}^{(c)}]^T$ and $\mv{s}_{l}$ = [$s_{1}^{(l)}$, $\cdots$, $s_{M_l}^{(c)}]^T$. 
					
					\item Estimate the total power of the interference given in (\ref{eq:MF effective channel model}), which can be equivalently expressed as $\Big(\mv{\tilde{h}}_{k,m}^{(c,c)}\Big)^H$ $\mv{W}_k^{(c)}$ $\mv{\tilde{h}}_{k,m}^{(c,c)}$, in which $\mv{W}_k^{(c)}$ is defined in (\ref{eq:INCM}); and compute the SINR corresponding to the $m$-th beam $\mv{\phi}_m^{(c)}$, which is expressed as
						\begin{align}\label{eq:SINR RBF-MF}
							\text{SINR}_{k,m}^{(\text{MF},c)} = \frac{\frac{P_T}{M_c}||\mv{\tilde{h}}_{k,m}^{(c,c)}||^4}
							 {\Big(\mv{\tilde{h}}_{k,m}^{(c,c)}\Big)^H \mv{W}_k^{(c)} 	\mv{\tilde{h}}_{k,m}^{(c,c)}}.
						\end{align}
					\item Send back $\text{SINR}_{k,m}^{(\text{MF},c)}$, $1\leq m\leq M_c$, to the $c$-th BS.
				\end{enumerate}
				
		\item[b3)] RBF-AS: The received signal at the $n$-th receive antenna of user $k$ in the $c$-th cell is given by	
			\begin{align}
				y_{k,n}^{(c)} = \mv{h}_{k,n}^{(c,c)}\sum_{m=1}^{M_c} \mv{\phi}_{m}^{(c)} {\it s}_m^{(c)} +\sum_{l=1,~l\neq c}^{C} \sqrt{\gamma_{l,c}} \mv{h}_{k,n}^{(l,c)}\sum_{m=1}^{M_l} \mv{\phi}_{m}^{(l)} {\it s}_m^{(l)} + z_{k,n}^{(c)}, 
			\end{align}
			for $1\leq n\leq N_R$, where $y_{k,n}^{(c)}$ and $z_{k,n}^{(c)}$ are the $n$-th element of $\mv{y}_{k}^{(c)}$ and $\mv{z}_{k}^{(c)}$, respectively; $\mv{h}_{k,n}^{(l,c)}$ $\in \mathbb{C}^{1\times M_l}$ is the $n$-th row of $\mv{H}_{k}^{(l,c)}$, $n$ $\in$ $\{1, \dots, N_R \}$, $l, c$ $\in$ $\{1, \dots, C\}$. For each of the $M_c$ beams, user $k$ does the following:
			\begin{enumerate}
				\item Estimate the SINR corresponding to the $m$-th beam $\mv{\phi}_{m}^{(c)}$ at the $n$-th antenna:		
				\begin{align}\label{eq:n-th SINR RBF-AS}
					\text{SINR}_{k,n,m} = \displaystyle \frac{\frac{P_T}{M_c}\left| \mv{h}_{k,n}^{(c,c)} \mv{\phi}_{m}^{(c)}\right|^2} { \displaystyle\frac{P_T}{M_c}\displaystyle\sum_{i=1,i\neq m}^{M_c} \left| \mv{h}_{k,n}^{(c,c)} \mv{\phi}_{i}^{(c)}\right|^2 +\displaystyle\sum_{l=1,l\neq c}^{C}\gamma_{l,c}\frac{P_T}{M_l} \displaystyle\sum_{i=1}^{M_l} \left| \mv{h}_{k,n}^{(l,c)} \mv{\phi}_{i}^{(l)}\right|^2 + \sigma^2 }.
				\end{align}
    	
				\item Select the antenna that has the largest SINR among all $N_R$ receive antennas for the $m$-th beam, and obtain the SINR as 
    			\begin{align}\label{eq:SINR RBF-AS}
    				\text{SINR}_{k,m}^{(\text{AS},c)} := \max_{n\in \{1, \cdots,N_R\}} \text{SINR}_{k,n,m}.
   			 	\end{align} 
    			\item Send back $\text{SINR}_{k,m}^{(\text{AS},c)}$, $1\leq m\leq M_c$, to the $c$-th BS.
    		\end{enumerate}
		\end{enumerate}	
	\item Transmission phase: After receiving the SINR feedback from all $K_c$ users, the $c$-th BS assigns the $m$-th beam to the user with the highest SINR for transmission, i.e.,
		\begin{align}
        	k_m^{(\text{Rx},c)}=\text{arg}\max_{k\in \{1, \cdots,K_c\}}\text{SINR}_{k,m}^{(\text{Rx},c)},
    	\end{align}
	where ``Rx'' denotes MMSE, MF, or AS.
		
\end{enumerate}

The achievable sum-rate in bits per second per Hz (bps/Hz) of the $c$-th cell by the above RBF scheme with different spatial receivers is then expressed as
\begin{align}\label{eq:c-th sum rate}
R_{\text{RBF-Rx}}^{(c)} = \mathbb{E}\left[\displaystyle\sum_{m=1}^{M_c}\log_2\left( 1+ \text{SINR}_{k_m^{(\text{Rx},c)},m}^{(\text{Rx},c)} \right) \right] \stackrel{(a)}{=} M_c\mathbb{E}\left[\log_2\left(1+\text{SINR}_{k_1^{(\text{Rx},c)},1}^{(\text{Rx},c)}\right)\right],
\end{align}
where $(a)$ is due to the fact that all the beams in each cell have the same SINR distribution with a given spatial receiver scheme.

\subsection{DoF Region}\label{subsec:DoF definitions}

In this paper, we apply the high-SNR analysis to draw insightful comparisons on the achievable rates of multi-cell MIMO RBF with different spatial diversity receivers. Similar to \cite{Hieu01}, we adopt the DoF region as one key performance metric in our analysis, which is defined as follows.

\begin{definition}\label{def:def. DoF region}{\it (General DoF region)}
The DoF region of a $C$-cell MIMO downlink system is defined as \cite{Jafar01}
\begin{align}\label{eq:def. DoF region}
		\mathcal{D}_{\text{MIMO}} = 
        \bigg\{ (d_1,\cdots,d_C)\in\mathbb{R}_{+}^{C}: \forall
        (\omega_1,\omega_2,\cdots,\omega_C)\in\mathbb{R}_{+}^{C}  ; ~~ \sum_{c=1}^{C}{\omega_cd_c}\leq \displaystyle\lim_{\rho\to\infty} \displaystyle\sup_{\mv{R}\in \mathcal{R}} \sum_{c=1}^{C}{\omega_c\frac{R_{sum}^{(c)}}{\log_2\rho}} \bigg\},
\end{align} 
where $\rho$ is the per-cell SNR; $\omega_c$, $d_c$, and $R_{sum}^{(c)}$ are the non-negative weight, achievable DoF, and sum rate of the $c$-th cell, respectively; and the region $\mathcal{R}$ is the set of all the achievable sum-rate tuples for all the cells, denoted by $\mv{R}=(R_{sum}^{(1)}, R_{sum}^{(2)}, \cdots, R_{sum}^{(C)})$.
\end{definition}

With RBF, the achievable DoF region in (\ref{eq:def. DoF region}) is more specifically given as follows.

\begin{definition}\label{def:def. DoF region RBF}{\it (DoF region with RBF)}
The DoF region of a $C$-cell MIMO downlink system with RBF is given by
	\begin{align}\label{eq:def. DoF region_RBF}
		\mathcal{D}_{\text{RBF-Rx}} 
		& = \bigg\{ (d_1,\cdots,d_C)\in\mathbb{R}_{+}^{C}:\forall
		(\omega_1,\omega_2,\cdots,\omega_C)\in\mathbb{R}_{+}^{C}  ; \notag \\
		& \qquad \qquad \qquad \sum_{c=1}^{C}{\omega_cd_c}\leq \displaystyle\lim_{\rho\to\infty} \bigg[
		\displaystyle\max_{M_1,\dots,M_C \in \{ 0, \cdots, N_T\}}
		\sum_{c=1}^{C}{\omega_c\frac{R_{\text{RBF-Rx}}^{(c)}}{\log_2\rho}}
		\bigg] \bigg\}.	
	\end{align}
where ``Rx'' denotes MMSE, MF or AS.
\end{definition}

Certainly, $\mathcal{D}_{\text{RBF-Rx}}$ $\subseteq$ $\mathcal{D}_{\text{MIMO}}$ regardless of MMSE, MF or AS spatial receivers used.

The high-SNR analysis preserves the interference-limited nature of a multi-cell system. However, in the case of RBF, we should also take into account the opportunistic user scheduling with sufficiently large number of users in each cell. To gain insight on the interplay between interference and multi-user diversity, it is practically useful to assume a certain growth rate for the number of users in each cell $K_c$ with respect to the SNR, $\rho$, as $\rho$ goes to infinity. Similar to \cite{Hieu01}, we make the following assumption. 

\begin{assumption}\label{assump:K with rho}
The number of users in the $c$-th cell scales with the SNR $\rho$ in the order of $\rho^{\alpha_c}$, $c=1, \dots, C$, with $\alpha_c\geq 0$, denoted by $K_c=\Theta(\rho^{\alpha_c})$, i.e., $K_c/\rho^{\alpha_c}\to a_c$ as $\rho\to\infty$ with $a_c$ being a positive constant independent of $\alpha_c$.
\end{assumption}

Here, $\alpha_c$ can be interpreted as a measure of the user density in the $c$-th cell; given the same coverage area for all cells, a larger $\alpha_c$ thus indicates more users in the $c$-th cell. We can consider the DoF region characterization under Assumption \ref{assump:K with rho} as an extension of the conventional approach with finite number of users to asymptotically large number of users with the increasing SNR. As will be seen later in this paper, such a characterization provides new insights on the different effects of the number of per-cell users, transmit beams, and receive antennas on the achievable rate in multi-cell MIMO RBF. The notations $\mathcal{D}_{\text{MIMO}}(\mv{\alpha})$ and  $\mathcal{D}_{\text{RBF-Rx}}(\mv{\alpha})$ will be used in the sequel to denote the DoF regions under Assumption \ref{assump:K with rho} with $K_c=\Theta(\rho^{\alpha_c})$, $c$ = 1, $\cdots$, $C$, and $\mv{\alpha}$ = [$\alpha_1$, $\cdots$, $\alpha_C]^T$. It is worth noting that our high-SNR approach is along the same line of those recently reported in \cite{Tajer01}-\cite{Lee01}, where the authors obtain the achievable DoF of their studied systems assuming that the number of users/links scales in a certain polynomial order with the SNR as the SNR goes to infinity.

\section{SINR Distribution}\label{sec:SINR distributions}

To characterize the achievable rates of the proposed RBF schemes with different spatial receivers, it is necessary to investigate the receiver SINRs given in (\ref{eq:SINR RBF-MMSE}), (\ref{eq:SINR RBF-MF}), and (\ref{eq:SINR RBF-AS}). In this section, we derive the (exact) distribution of the SINR in each receiver case. 

\subsection{RBF-MMSE}\label{subsec:SINR RBF-MMSE}

To obtain the SINR distribution for the RBF-MMSE scheme, we first prove a more general result in MVA, which is given as follows.

\begin{theorem}\label{theorem:general}
Given $\mv{h}\sim\mathcal{CN}(\mv{0}_{p\times 1},\mv{I}_p)$, $\mv{X}\sim\mathcal{CN}(\mv{0}_{p\times n},\mv{I}_p\otimes\mv{I}_n)$\footnotemark \footnotetext{\mv{X} is said to have a matrix-variate complex Gaussian distribution with mean matrix $\mv{0}$ $\in \mathbb{C}^{p \times n}$ and covariance matrix $\mv{I}_p\otimes\mv{I}_n$, where $\otimes$ denotes the Kronecker product.}, $n\geq p\geq 1$, where $\mv{h}$ is independent of $\mv{X}$, and $\mv{\Psi}$ = $diag$($\psi_1$, $\ldots$, $\psi_n$), with $\psi_i > 0$, $i=1,\ldots,n$, being constants, the cumulative distribution function (CDF) of the random variable $S := \mv{h}^H(\mv{X}\mv{\Psi}\mv{X}^H)^{-1}\mv{h}$ is given by
	\begin{align}\label{eq:general} 
		F_{S}(s) = \frac{\sum_{i=p}^{n}\beta_i s^i}{\prod_{i=1}^n (1+\psi_i s)},
	\end{align}
where $\beta_i$ is the coefficient of $s^i$ after expanding the polynomial $\prod_{j=1}^n (1+\psi_j s)$.	
\end{theorem}
\begin{IEEEproof}
	Please refer to Appendix \ref{proof:general}.\footnotemark
\end{IEEEproof}
\footnotetext{Note that a similar result of Theorem \ref{theorem:general} has been reported in \cite{Gao}, but via a different proof method. Specifically, the authors in \cite{Gao} applied a ``top-down'' approach, whereby they used a more general result \cite[Theorem 3 and (59)]{Khatri} to derive the explicit expression (\ref{eq:general}) for the case in Theorem \ref{theorem:general}. In this paper, we propose an alternative more direct approach, which uses only fundamental properties in MVA and thus leads to a more compact proof.}

It is worth noting that extensions of Theorem \ref{theorem:general} to the case of Rician-fading and/or correlated channels can be found in subsequent studies, e.g., \cite{Shah01}-\cite{McKay01}, where the moment generating function and distribution of the output SINR have been derived. These results are then applied to find the closed-form expressions of the capacity and/or bit-error-rate for the investigated systems. Under such cases, the SINR distribution in general possesses a complicated form and is often expressed in terms of determinants of certain matrices. 

Next, we observe that (\ref{eq:general}) can be equivalently expressed as
\begin{align}\label{eq:general 2} 
	F_{S}(s) = 1 - \frac{\sum_{i=0}^{p-1}\beta_i s^i}{\prod_{i=1}^n (1+\psi_i s)}.
\end{align}

We are now ready to obtain the SINR distribution with RBF-MMSE, based on Theorem \ref{theorem:general}.
\begin{corollary}\label{corollary:SINR RBF-MMSE CDF}
Given $N_R\leq \sum_{l=1}^{C}M_l - 1$, the CDF of the random variable $S := \text{SINR}_{k,m}^{(\text{MMSE},c)}$ defined in (\ref{eq:SINR RBF-MMSE}) is given by
	\begin{align}\label{eq:SINR RBF-MMSE CDF}
		F_{S}(s) = 1 - \frac{e^{-s/\eta_c} \left( \sum_{i=0}^{N_R - 1}\zeta_i s^i \right) }{(1+ s)^{M_c-1} \prod_{l=1, l\neq c}^{\sum M_c - 1} (1+ \frac{\mu_{l,c}}{\eta_c} s)^{M_l}},
	\end{align}
where $\zeta_i$ is the coefficient of $s^i$ in the polynomial expansion of $(1+ s)^{M_c-1} \prod_{l=1, l\neq c}^{\sum M_c - 1} (1+ \frac{\mu_{l,c}}{\eta_c} s)^{M_l}$.	
\end{corollary}
\begin{IEEEproof}
	Please refer to Appendix \ref{proof:SINR RBF-MMSE CDF}.
\end{IEEEproof}

\subsection{RBF-MF}\label{subsec:SINR RBF-MF}
		
The interference-plus-noise covariance matrix $\mv{W}_k^{(c)}$ given in (\ref{eq:INCM}) can be alternatively expressed as
\begin{align}
	\mv{W}_k^{(c)} = \Big(\mv{\tilde{H}}_{k,m}^{(c)}\Big)^H  diag \bigg( \underbrace{\frac{P_T}{M_c}, \cdots, \frac{P_T}{M_c}}_{M_c-1}, \cdots, \underbrace{\frac{P_T\gamma_{l,c}}{M_l}, \cdots, \frac{P_T\gamma_{l,c}}{M_l}}_{M_l}, \cdots \bigg) \mv{\tilde{H}}_{k,m}^{(c)} + \sigma^2\mv{I},
\end{align}
where $\mv{\tilde{H}}_{k,m}^{(c)} = \left[ \mv{\tilde{H}}_{k,-m}^{(c,c)}, \mv{\tilde{H}}_{k}^{(1,c)}, \cdots, \mv{\tilde{H}}_{k}^{(l,c)}, \cdots, \mv{\tilde{H}}_{k}^{(C,c)} \right]$; $\mv{\tilde{H}}_{k,-m}^{(c,c)}$ and $\mv{\tilde{H}}_{k}^{(l,c)}$ are defined in (\ref{eq:INCM}), $l, c \in \{ 1, \dots, C\}$, $l\neq c$. Define
\begin{align}
	\mv{\hat{h}}_{k,m}^{(c)} = \frac{\mv{\tilde{h}}_{k,m}^{(c,c)}}{\big\|\mv{\tilde{h}}_{k,m}^{(c,c)}\big\|} \mv{\tilde{H}}_{k,m}^{(c)}.
\end{align} 

Note that $\mv{\hat{h}}_{k,m}^{(c)}$ $\in$ $\mathbb{C}^{(\sum_{l=1}^C M_l - 1) \times 1}$ is independent of $\mv{\tilde{h}}_{k,m}^{(c,c)}$ and all the elements of $\mv{\hat{h}}_{k,m}^{(c)}$ are i.i.d. CSCG random variables each distributed as $\mathcal{CN}(0,1)$. For RBF-MF, the SINR in (\ref{eq:SINR RBF-MF}) is thus expressed as
\begin{align}\label{eq:SINR RBF-MF 2} 
	\text{SINR}_{k,m}^{(\text{MF},c)} = \frac {||\mv{\tilde{h}}_{k,m}^{(c,c)}||^2} 
	{ \left(\mv{\hat{h}}_{k,m}^{(c)}\right)^H \mv{G} \mv{\hat{h}}_{k,m}^{(c)} + \frac{1}{\eta_c} },
\end{align}
where $\mv{G} = diag \bigg( \underbrace{1, \cdots, 1}_{M_c-1}, \cdots, \underbrace{ \frac{\mu_{l,c}}{\eta_c}, \cdots, \frac{\mu_{l,c}}{\eta_c} }_{M_l}, \cdots \bigg)$, with $l, c \in \{ 1, \dots, C\}$, $l\neq c$. 

To the authors' knowledge, there is no distribution result in the literature regarding the random variable with the form (\ref{eq:SINR RBF-MF 2}). By applying the \emph{characteristic function} approach, we obtain the CDF of the SINR with RBF-MF in the following theorem.

\begin{theorem}\label{theorem:SINR RBF-MF CDF}
	The CDF of the random variable $S := \text{SINR}_{k,m}^{(\text{MF},c)}$ in (\ref{eq:SINR RBF-MF 2}) is given by
	\begin{align}\label{eq:SINR RBF-MF CDF}
		F_S(s) = 1 - e^{-s/\eta_c}\displaystyle\sum_{k=0}^{N_R-1}\sum_{m=0}^k \frac{(-1)^m s^k} 
		{ (k-m)! m!  \eta_c^{k-m}} \frac{d^mT_0(s)}{ds^m},
	\end{align} 
	where
	\begin{align}\label{eq:T_0}
		T_0(s) = \frac{1}{(1+s)^{M_c-1}\prod_{l=1,l\neq c}^{C}{(1+\frac{\mu_{l,c}}{\eta_c}s)^{M_l} }}.
	\end{align}
\end{theorem}
\begin{IEEEproof}
	Please refer to Appendix \ref{proof:SINR RBF-MF CDF}.	
\end{IEEEproof}			
 
\subsection{RBF-AS}\label{subsec:SINR RBF-AS}

First, we investigate the distribution of the $\text{SINR}_{k,n,m}$ given in (\ref{eq:n-th SINR RBF-AS}). Note that $ \left| \mv{h}_{k,n}^{(l,c)} \mv{\phi}_{i}^{(l)} \right|^2$, $\forall k, n, l, c, i$, are i.i.d. chi-square random variables with 2 degrees of freedom, denoted by $\chi^2(2)$ \cite{Sharif01}. From Corollary \ref{corollary:SINR RBF-MMSE CDF} or Theorem \ref{theorem:SINR RBF-MF CDF}, we can easily obtain the same distribution for $\text{SINR}_{k,n,m}$ and thereby $\text{SINR}_{k,m}^{(\text{AS},c)}$ in (\ref{eq:SINR RBF-AS}), as given in the following corollary.

\begin{corollary}\label{coro:SINR RBF-AS CDF}
	The CDF of the random variable $S := \text{SINR}_{k,m}^{(\text{AS},c)}$ defined in (\ref{eq:SINR RBF-AS}) is given by
	\begin{align}\label{eq:SINR RBF-AS CDF}
		F_S(s) = \Vast( 1-\frac{e^{-s/\eta_c}}{{\left(s+1\right)}^{M_c-1}
            \displaystyle\prod_{l=1,l\neq c}^C{\left(\frac{\mu_{l,c}}{\eta_c}s+1\right)^{M_l}}} \Vast)^{N_R}.
	\end{align} 
\end{corollary}

\begin{figure}[t]
    \centering
    \epsfxsize=0.65\linewidth
  	\includegraphics[width=11.5cm, height=8.5cm]{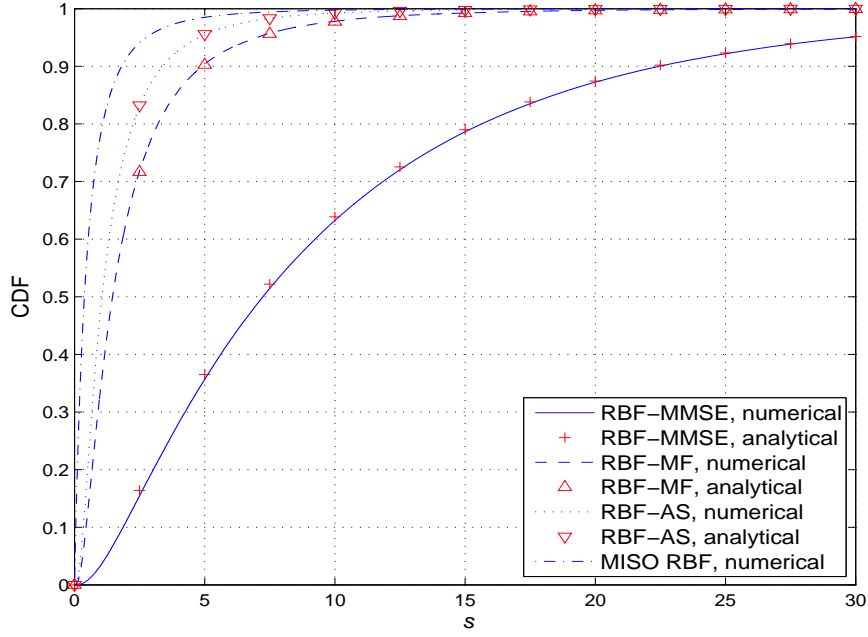}
		\captionsetup{width=0.65\textwidth}
    \caption{Comparison of the simulated and analytical CDFs of the SINR with different spatial receiver schemes.}\label{fig:1}
    \vspace{-0.15in}
\end{figure}

In Fig. \ref{fig:1}, we show the SINR CDFs of the RBF-MMSE, RBF-MF, and RBF-AS schemes under the following setup: $C$ = 4, $\eta_1$ = 20 dB, $N_R$ = 3, $M_1$ = 3, [$\mu_{2,1}$, $\mu_{3,1}$, $\mu_{4,1}$] = [0, $-3$, 3] dB, and [$M_2$, $M_3$, $M_4$] = [3, 2, 4]. The CDFs obtained by Monte-Carlo simulations are compared to our analytical results from Corollary \ref{corollary:SINR RBF-MMSE CDF}, Theorem \ref{theorem:SINR RBF-MF CDF}, and Corollary \ref{coro:SINR RBF-AS CDF}. It is observed that both analytical and simulation results match closely. For comparison, we also plot the SINR CDF in the case with $N_R=1$, i.e., the MISO RBF scheme that was investigated in \cite{Hieu01}. It is observed that receive spatial diversity helps enhance the SINR performance substantially. In particular, with RBF-MMSE, the SINR distribution is most significantly improved. It is thus expected that RBF-MMSE should also provide the best rate performance, as will be shown next.

\section{DoF Analysis}\label{sec:DoF characterization}

In this section, we first study the DoF for the achievable sum-rate in the single-cell MIMO RBF case. Then, we extend the DoF analysis for the single-cell RBF to the more general multi-cell RBF subject to the inter-cell interference. Finally, we investigate the optimality of RBF in terms of achievable DoF region.

\subsection{Single-Cell Case}
First, we consider the single-cell case without the inter-cell interference to draw some useful insights. For brevity, we drop the cell index $c$ in this subsection. We define the achievable DoF for the sum-rate in one single cell with a given pair of user density $\alpha$ and number of transmit beams $M$ as
\begin{align}
d_{\text{RBF-Rx}}(\alpha,M) = \lim_{\rho\to\infty}\frac{R_{\text{RBF-Rx}}}{\log_2 \rho}.
\end{align} 

We first obtain the following lemma on the achievable DoF in one single cell.

\begin{lemma}\label{lemma:theoDoF single-cell RBF-MMSE, RBF-MF/AS}
In the single-cell case, given $K=\Theta\left(\rho^{\alpha}\right)$, the achievable DoF of RBF-MMSE, RBF-MF, and RBF-AS schemes are given by
\begin{subnumcases}{d_{\text{RBF-MMSE}}(\alpha,M) =}
    \frac{\alpha M}{M-N_R}, \hspace{-0.2in} & $0\leq \alpha\leq M-N_R$ \hspace{0.2in} \label{case:theoDoF single-cell RBF-MMSE a} \\
    M, \hspace{-0.2in} & $\alpha > M-N_R$. \label{case:theoDoF single-cell RBF-MMSE b}
\end{subnumcases}
\begin{subnumcases}{d_{\text{RBF-MF/AS}}(\alpha,M) =}
    \frac{\alpha M}{M-1}, \hspace{-0.2in} & $0\leq \alpha\leq M-1$ \hspace{0.2in} \label{case:theoDoF single-cell RBF-MF/AS a} \\
    M, \hspace{-0.2in} & $\alpha > M-1$. \label{case:theoDoF single-cell RBF-MF/AS b}
\end{subnumcases}
\end{lemma}
\begin{IEEEproof}
Please refer to Appendix \ref{proof:theoDoF single-cell RBF-MMSE, RBF-MF/AS}.
\end{IEEEproof}

\begin{figure}[t]
    \centering
    \epsfxsize=0.65\linewidth
  	\includegraphics[width=11.5cm, height=7.5cm]{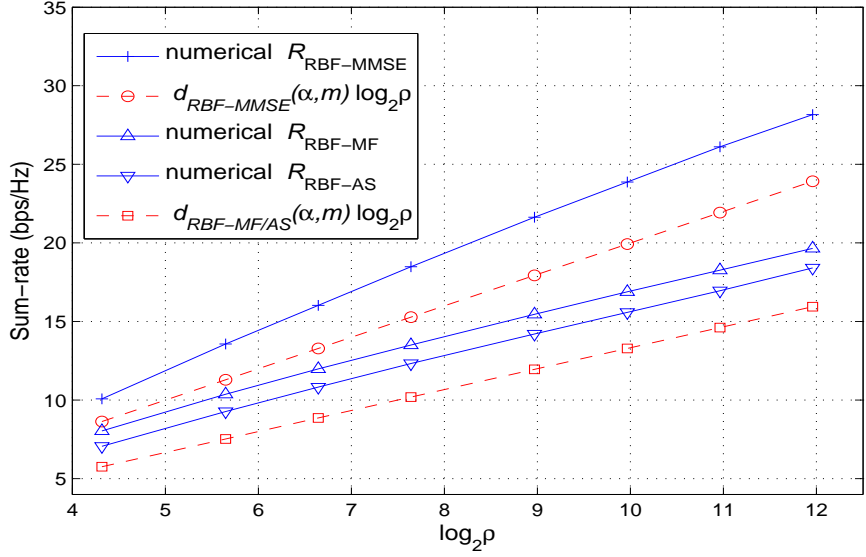}
  	\vspace{-0.1in}
  	\captionsetup{width=0.65\textwidth}
    \caption{Comparison of the numerical sum-rate and sum-rate scaling law in the single-cell MIMO RBF with different spatial receivers.}
    \label{fig:DoF scaling law}
    \vspace{-0.3in}
\end{figure}

In Fig. \ref{fig:DoF scaling law}, the sum-rate scaling laws $d_{\text{RBF-MMSE}}(\alpha,M) \log_2\rho$ and $d_{\text{RBF-MF/AS}}(\alpha,M) \log_2\rho$ are compared with the actual sum-rates achievable by RBF-MMSE, RBF-MF, and RBF-AS obtained by simulation. The system parameters are set as $M$ = 4, $N_R$ = 2, $\alpha$ = 1, and $K$ = $\lfloor\rho^{\alpha}\rfloor$. A good match between the theoretical rate scaling laws and numerical sum-rate results is observed, even with moderate SNR values of $\rho$. 

From Lemma \ref{lemma:theoDoF single-cell RBF-MMSE, RBF-MF/AS}, we observe an interesting interplay among the available multi-user diversity (specified by the user density $\alpha$), the level of the intra-cell interference (specified by $M-1$), the receive diversity gain (specified by the number of receive antennas $N_R$), and the achievable spatial multiplexing gain (specified by the DoF $d_{RBF}(\alpha,M)$), which is elaborated as follows. 

First, note that in a single-cell RBF system, transmitting $M$ beams simultaneously results in $M-1$ intra-cell interfering beams for each received beam. The term $M-1$ in the denominator of (\ref{case:theoDoF single-cell RBF-MF/AS a}) is exactly the number of interfering beams to one particular received beam in RBF-MF/AS. However, there exist only $M-N_R$ effective interference beams in RBF-MMSE, as shown in (\ref{case:theoDoF single-cell RBF-MMSE a}), since MMSE receiver achieves an additional \emph{interference mitigation} gain of $N_R-1$. Specifically, with the total $N_R$ spatial DoF, the MMSE receiver effectively uses one DoF for receiving signal and the other $N_R-1$ DoF for suppressing the interference. Furthermore, in terms of achievable sum-rate DoF, the performance of either RBF-MF or RBF-AS is the same as that of MISO RBF system without receive spatial diversity \cite{Hieu01}, and is thus poorer as compared to RBF-MMSE with $N_R >1$. The DoF gain by receive spatial diversity therefore clearly depends on the availability of the interference covariance matrix at each MS. In the case of RBF-MMSE, the interference-plus-noise covariance matrix $\mv{W}_k^{(c)}$ in (\ref{eq:INCM}) needs to be estimated at the receiver, while this operation is not required in RBF-MF or RBF-AS. 

Another interpretation of Lemma \ref{lemma:theoDoF single-cell RBF-MMSE, RBF-MF/AS} is that it gives the \emph{user scaling law} with SNR required to achieve $d$ DoF, similarly to \cite{Tajer01}-\cite{Lee01}. Specifically, the number of users should scale as $K$ = $\Theta\left(\rho^{d\frac{M-N_R}{M}} \right)$ and $\Theta\left(\rho^{d\frac{M-1}{M}} \right)$ for RBF-MMSE and RBF-MF/AS, respectively. Thus, significantly less number of users is required in RBF-MMSE as compared to RBF-MF/AS for achieving the same DoF. With RBF and under the assumption $K = \Theta(\rho^{\alpha})$, it is also interesting to observe from Lemma \ref{lemma:theoDoF single-cell RBF-MMSE, RBF-MF/AS} that the achievable DoF can be a non-negative real number (as compared to the conventional integer DoF in the literature with finite $K$). This comes from our (quite general) assumption that $\alpha$ can take any arbitrary real value.

Next, we obtain the maximum achievable DoF of RBF-MMSE for a given $\alpha$ by searching over all possible values of $M$.We note that for any $M$ 
$<$ $\lfloor\alpha\rfloor+N_R$, $d_{RBF}(\alpha,M)$ $<$ 
$d_{RBF}(\alpha,\lfloor\alpha\rfloor+N_R)$, while for any $M$ $>$ 
$\lfloor\alpha\rfloor+N_R+1$, $d_{RBF}(\alpha,M)$ $<$ 
$d_{RBF}(\alpha,\lfloor\alpha\rfloor+N_R+1)$. Thus we only need to compare 
$d_{RBF}(\alpha,\lfloor\alpha\rfloor+N_R)$ and $d_{RBF}(\alpha,\lfloor\alpha\rfloor+N_R+1)$ in searching for the optimal $M$. The maximum achievable DoF of RBF-MF/AS can be obtained similarly. The result is shown in the following theorem.

\begin{theorem}\label{theorem:maximum DoF single-cell}
For a single-cell MIMO RBF system with $N_T$ transmit antennas, $N_R$ receive antennas, and user density coefficient $\alpha$, the maximum achievable DoF and the corresponding optimal number of transmit beams with MMSE, MF, or AS based receivers are\footnotemark \footnotetext{The notations $\lfloor\alpha\rfloor$ and \{$\alpha$\} denote the integer and fractional parts of a real number $\alpha$, respectively.} 
\begin{align}\label{eq:RBF-MMSE single dstar}
        d_{\text{RBF-MMSE}}^*(\alpha) &=
            \begin{cases}
            \lfloor\alpha\rfloor+N_R, & \alpha\leq N_T-N_R, N_R\geq \{\alpha\}(\lfloor\alpha\rfloor+N_R+1), \\
            \frac{\alpha(\lfloor\alpha\rfloor+N_R+1)}{\lfloor\alpha\rfloor+1},
            & \alpha\leq N_T-N_R, \{\alpha\}(\lfloor\alpha\rfloor+N_R+1)>N_R, \\
            N_T, & \alpha> N_T-N_R.
            \end{cases} \\ {} \label{eq:RBF-MMSE single Mstar}
        M_{\text{RBF-MMSE}}^*(\alpha) &=
            \begin{cases}
            \lfloor\alpha\rfloor+N_R, & \alpha\leq N_T-N_R, N_R\geq \{\alpha\}(\lfloor\alpha\rfloor+N_R+1), \\
            \lfloor\alpha\rfloor+N_R+1, & \alpha\leq N_T-N_R, \{\alpha\}(\lfloor\alpha\rfloor+N_R+1)>N_R, \\
            N_T, & \alpha> N_T-N_R.
            \end{cases}
\end{align}
\begin{align}\label{eq:RBF-MF/AS single dstar}
        d_{\text{RBF-MF/AS}}^*(\alpha) &=
            \begin{cases}
            \lfloor\alpha\rfloor+1, & \alpha\leq N_T-1, 1\geq \{\alpha\}(\lfloor\alpha\rfloor+2), \\
            \frac{\alpha(\lfloor\alpha\rfloor+2)}{\lfloor\alpha\rfloor+1},
            & \alpha\leq N_T-1, \{\alpha\}(\lfloor\alpha\rfloor+2)>1, \\
            N_T, & \alpha> N_T-1.
            \end{cases}\\{}
        \label{eq:RBF-MF/AS single Mstar}
        M_{\text{RBF-MF/AS}}^*(\alpha) &=
            \begin{cases}
            \lfloor\alpha\rfloor+1, & \alpha\leq N_T-1, 1\geq \{\alpha\}(\lfloor\alpha\rfloor+2), \\
            \lfloor\alpha\rfloor+2, & \alpha\leq N_T-1, \{\alpha\}(\lfloor\alpha\rfloor+2)>1, \\
            N_T, & \alpha> N_T-1.
            \end{cases}
\end{align}
\end{theorem}

\begin{figure}
    \centering
    \epsfxsize=0.65\linewidth
    \captionsetup{width=0.65\textwidth}
  	\includegraphics[width=11.5cm, height=8.5cm]{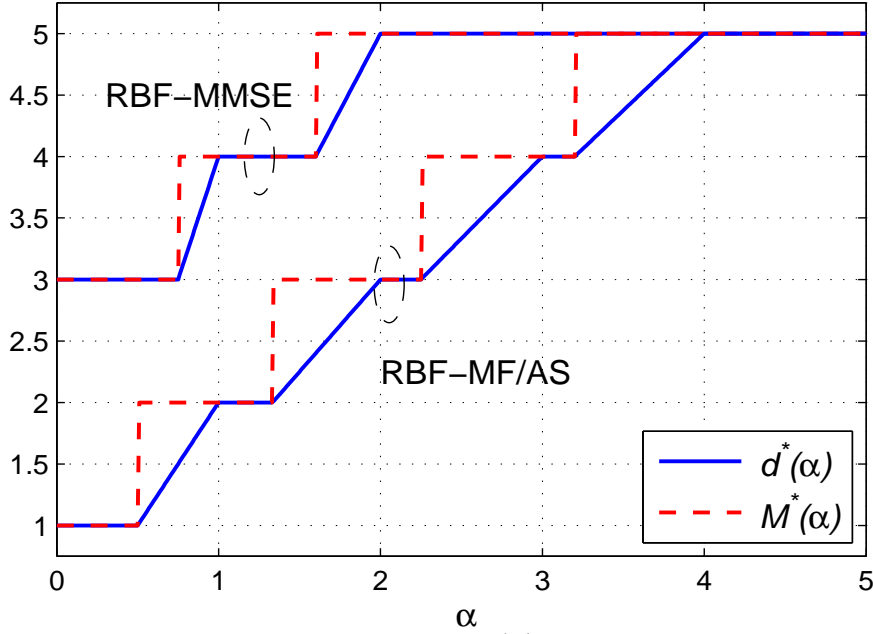}
  	\vspace{-0.2in}
    \caption{The maximum sum-rate DoF $d_{\text{RBF-Rx}}^*(\alpha)$ and optimal number of transmit beams $M_{\text{RBF-Rx}}^*(\alpha)$ with $N_T=5$ and $N_R=3$, where ``Rx'' denotes MMSE, MF, or AS. }
    \label{fig:DoFsingle}
    \vspace{-0.2in}
\end{figure}

In Fig. \ref{fig:DoFsingle}, we show the maximum DoF and the corresponding optimal number of transmit beams versus the user density coefficient $\alpha$ with $N_T=5$ and $N_R=3$ for each single-cell RBF scheme, according to Theorem \ref{theorem:maximum DoF single-cell}. It is observed that in general, RBF-MMSE achieves a higher maximum DoF by transmitting more data beams as compared to RBF-MF or RBF-AS. As a result, RBF-MMSE system can serve more users with better rate performance than RBF-MF/AS. However, the improvement in the achievable rate and coverage comes at the cost of higher complexity by employing MMSE receivers.
One important question is how the RBF schemes perform as compared to the optimal DPC-based transmission scheme assuming the full transmitter-side CSI in single-cell MIMO BCs. In the following, we answer this question in terms of achievable sum-rate DoF. First, we obtain an upper bound on the single-cell achievable DoF with arbitrary transmission schemes.

\begin{proposition}\label{prop:single-cell DoF upperbound}
Assuming $K=\Theta(\rho^{\alpha})$ with $\alpha \geq 0$, the DoF of a single-cell MIMO BC with $N_T$ transmit antennas at the BS and $N_R$ receive antennas at each MS is upper-bounded by $N_T$ as $\rho\to\infty$.
\end{proposition}
\begin{IEEEproof}
Please refer to Appendix \ref{proof:single-cell DoF upperbound}.
\end{IEEEproof}

Proposition \ref{prop:single-cell DoF upperbound} states that the maximum DoF of the single-cell MIMO BC is always $N_T$, even with asymptotically large number of users that scales with the increasing SNR. Next, applying Theorem \ref{theorem:maximum DoF single-cell} and Proposition \ref{prop:single-cell DoF upperbound} yields the following proposition.

\begin{proposition}\label{prop:optimality single-cell}
Assuming $K=\Theta(\rho^{\alpha})$, the single-cell RBF schemes are DoF-optimal, i.e., $d_{\text{RBF-MMSE}}^*(\alpha)$ = $N_T$ and $d_{\text{RBF-MF/AS}}^*(\alpha)$ = $N_T$, if and only if
\begin{itemize}
\item RBF-MMSE: $\alpha \geq N_T-N_R$;
\item RBF-MF/AS: $\alpha \geq N_T-1$.
\end{itemize} 
\end{proposition}

It thus follows that the single-cell RBF schemes achieve the maximum DoF with $M=N_T$ if the number of users per-cell is sufficiently large, thanks to the multiuser diversity and/or spatial diversity that completely eliminates the intra-cell interference. However, spatial diversity gain in the achievable DoF is available only in the case of MMSE based receiver.  

\begin{figure}[t]
    \centering
    \epsfxsize=0.65\linewidth
    \captionsetup{width=0.85\textwidth}
    \includegraphics[width=11.5cm, height=8.5cm]{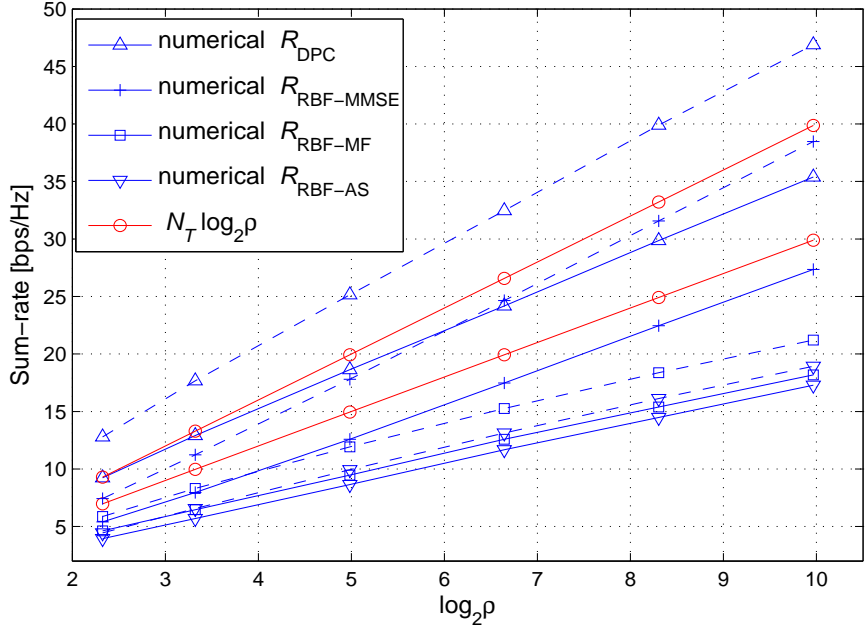}
  	\vspace{-0.1in}
    \caption{Comparison of the numerical DPC, RBF-MMSE, RBF-MF, and RBF-AS sum-rates, and the DoF scaling law with $N_T-1$ $\geq$ $\alpha \geq N_T-N_R$.}
    \label{fig:optimal DoF scaling law}
    \vspace{-0.2in}
\end{figure}

As an example for illustration, we compare the numerical sum-rates and the DoF scaling law in Fig. \ref{fig:optimal DoF scaling law}, in which the DPC, RBF-MMSE, RBF-MF, and RBF-AS are employed, and $N_T-1$ $>$ $\alpha$ $\geq$ $N_T-N_R$. We consider two single-cell systems with the following parameters: (a) $M$ = $N_T$ = 3, $N_R$ = 2, $\alpha$ = 1, $K$ = $\lfloor \rho^{\alpha}\rfloor$; and (b) $M$ = $N_T$ = 4, $N_R$ = 3, $\alpha$ = 1.2, $K$ = $\lfloor \rho^{\alpha}\rfloor$. The rates and scaling law of system (a) and (b) are denoted as the solid and dash lines, respectively. Note that in both cases, the DPC and RBF-MMSE sum-rates follow the (same) DoF scaling law quite closely. This example clearly demonstrates the DoF optimality of the RBF-MMSE given that $\alpha \geq N_T-N_R$. Furthermore, since $\alpha < N_T-1$, the RBF-MF, RBF-AS, and consequently MISO RBF schemes are DoF sub-optimal as clearly shown in Fig. \ref{fig:optimal DoF scaling law}. It is important to note that the values of the SNR and the numbers of users are only moderate in this example. This thus shows the practical usefulness of our optimality conditions for RBF schemes given in Proposition \ref{prop:optimality single-cell}.

Next, we compare our new asymptotic result to the conventional one in \cite{Sharif01} and \cite{Sharif02} with finite per-cell SNR, which states that for any given $M\leq N_T$ and $N_R\geq 1$, the sum-rate achievable by single-cell RBF satisfies
\begin{align}\label{eq:K to infinity}
\displaystyle\lim_{K\to\infty}\frac{R_{\text{RBF-Rx}}}{M\log_2\log K}=1,
\end{align}
where ``Rx'' denotes any of MMSE, MF, and AS.

We observe a notable difference between the conclusions drawn from our high-SNR analysis and the conventional finite-SNR analysis both with the number of per-cell users increasing to infinity. In the finite-SNR case, from (\ref{eq:K to infinity}) it follows that there is no asymptotic sum-rate gain by RBF-MMSE over RBF-MF or RBF-AS, and the asymptotic sum-rate is independent of $N_R$. This thus leads to an improper conclusion that using only one single antenna at each receiver is sufficient to capture the asymptotic rate of RBF. As a result, the benefit of receive spatial diversity is neglected, which in turn severely degrades the RBF rate performance especially for interference-limited multi-cell systems. However, with our high-SNR analysis, the effects of the number of receive antennas as well as the spatial diversity technique used (MMSE versus MF/AS) on the DoF performance are clearly shown. This demonstrates the advantage of our new approach for designing practical multi-cell systems employing RBF.

\subsection{Multi-Cell Case}\label{sec:Multi-cell case}

In this subsection, we extend the DoF analysis for the single-cell case to the more general multi-cell RBF. For convenience, we define the achievable sum-rate DoF of the $c$-th cell as $d_{\text{RBF-Rx},c}(\alpha_c,\mv{m})$ = $\lim_{\rho\to\infty}\frac{R_{\text{RBF-Rx}}^{(c)}}{\log_2 \rho}$, where $\mv{m}$ = [$M_1$,$\cdots$,$M_C]^T$ is a given set of numbers of transmit beams at different BSs. We then state the following result on the achieve DoF of the $c$-th cell.

\begin{lemma}\label{lemma:theoDoF c-cell RBF-MMSE, RBF-MF/AS}
In the multi-cell case, given $K_c=\Theta\left(\rho^{\alpha_c}\right)$ and $\mv{m}$, the achievable DoF of the $c$-th cell with RBF-MMSE, RBF-MF, and RBF-AS schemes are given by
\begin{subnumcases}{d_{{\text{RBF-MMSE},c}}(\alpha_c,\mv{m})=}
    \frac{\alpha_c M_c}{\sum_{l=1}^{C}M_l-N_R}, \hspace{-0.2in} & $0\leq \alpha_c\leq \sum_{l=1}^{C}M_l-N_R$ \hspace{0.2in} \label{case:theoDoF c-cell RBF-MMSE a} \\
    M_c, & $\alpha_c > \sum_{l=1}^{C}M_l-N_R$. \label{case:theoDoF c-cell RBF-MMSE b}
\end{subnumcases}

\begin{subnumcases}{d_{{\text{RBF-MF/AS},c}}(\alpha_c,\mv{m})=}
    \frac{\alpha_c M_c}{\sum_{l=1}^{C}M_l-1}, \hspace{-0.2in} & $0\leq \alpha_c\leq \sum_{l=1}^{C}M_l-1$ \hspace{0.2in} \label{case:theoDoF c-cell RBF-MF/AS a} \\
    M_c, & $\alpha_c > \sum_{l=1}^{C}M_l-1$. \label{case:theoDoF c-cell RBF-MF/AS b}
\end{subnumcases}
\end{lemma}

The proof of the above lemma can be obtained by similar arguments as for Lemma \ref{lemma:theoDoF single-cell RBF-MMSE, RBF-MF/AS}, and is thus omitted for brevity. Compared to the single-cell case, in the multi-cell case there are not only $M_c-1$ intra-cell interfering beams, but also $\sum_{l=1,l\neq c}^{C}M_l$ inter-cell interfering beams for any received beam in the $c$-th cell, as observed from the denominators in (\ref{case:theoDoF c-cell RBF-MMSE a}) and (\ref{case:theoDoF c-cell RBF-MF/AS a}), which results in a decrease in the achievable DoF per cell. 

We again compare our new asymptotic result to that obtained from the conventional asymptotic analysis with finite per-cell SNR \cite{Sharif01}, \cite{Sharif02}. We first note the following result, which states that for any given $M_c\leq N_T$ and $N_R\geq 1$, the sum-rate achievable by the $c$-th cell RBF satisfies
\begin{align}\label{eq:multicell K to infinity}
\displaystyle\lim_{K_c\to\infty}\frac{R^{(c)}_{\text{RBF-Rx}}}{M_c\log_2\log K_c}=1,
\end{align}
where ``Rx'' denotes any of MMSE, MF, and AS. Thus (\ref{eq:K to infinity}) and (\ref{eq:multicell K to infinity}) imply that the rate performance of each cell with any given number of receive antennas at the users in a multi-cell RBF system is equivalent to that of a single-cell RBF with \emph{one antenna at each user}. Such a conclusion may be misleading in a practical multi-cell system with non-negligible ICI where receive spatial diversity can help significantly improve the rate performance based on our new DoF analysis. Furthermore, (\ref{eq:multicell K to infinity}) implies that even in multi-cell RBF systems, each BS should use \emph{all} available orthogonal beams for transmission, i.e., $M_c$ = $N_T$, $\forall$ $c$ = $1,\dots,C$. This conclusion can severely degrade the rate performance of RBF systems, as illustrated in the next example.

\begin{figure}
        \centering
        \begin{subfigure}[b]{0.5\textwidth}
                \centering
                \includegraphics[width=7.5cm,height=7.0cm]{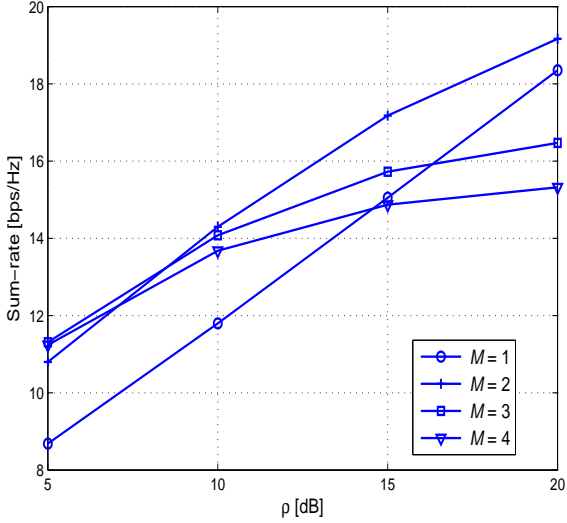}
                \caption{Two-cell}
                \label{fig:fig5}
        \end{subfigure}%
        ~ 
        \begin{subfigure}[b]{0.5\textwidth}
                \centering
                \includegraphics[width=7.5cm,height=7.0cm]{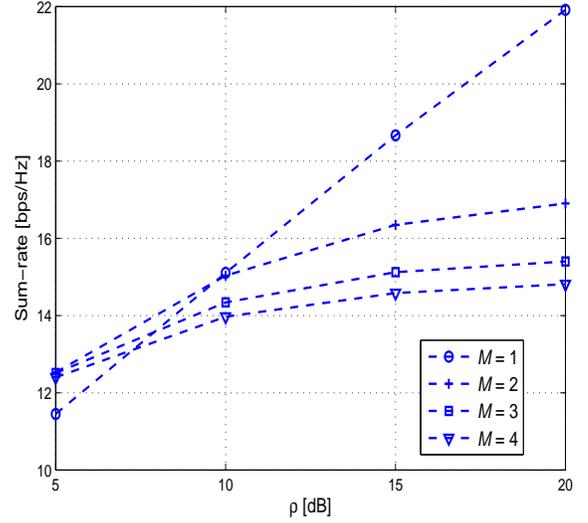}
                \caption{Three-cell}
                \label{fig:fig5b}
        \end{subfigure}
        \caption{Sum-rates of RBF-MMSE systems as a function of the SNR.}\label{fig:RBF-MMSE systems}
        \vspace{-0.2in}
\end{figure}

In Fig. \ref{fig:RBF-MMSE systems}, we depict the (total) achievable sum-rates of two RBF-MMSE systems with the following parameters: (a) $C$ = 2, $M_1$ = $M_2$ = $M$ $\leq$ $N_T=4$, $N_R=2$, $\gamma_{1,2}$ = $\gamma_{2,1}$ = 0.8, and $K_1$ = $K_2$ = $K$ = 200; and (b) $C$ = 3, $M_1$ = $M_2$ = $M_3$ = $M$ $\leq$ $N_T=4$, $N_R=2$, $\gamma_{l,c}$ = 0.8, $l$, $c$ = 1, 2, 3, $l\neq c$, and $K_1$ = $K_2$ = $K_3$ = $K$ = 200. Thus, with $\rho$ = [5 10 15 20] dB, we have $K$ $\approx$ $\lfloor \rho^{\alpha} \rfloor$, where $\alpha$ = [4.6021 2.3010 1.5340 1.1505]. Consider first system (a). From (\ref{eq:multicell K to infinity}), the conventional asymptotic analysis implies that the optimal rate performance is achieved with $M_1$ = $M_2$ = 4. However, given the constraint $M_1$ = $M_2$ = $M$, Lemma \ref{lemma:theoDoF c-cell RBF-MMSE, RBF-MF/AS} suggests that the best rate performance is achieved with $M$ = 3 when $\rho$ = 5 dB and $M$ = 2 for the other cases. The reason is that the discrete function $\frac{\alpha M}{2M-N_R}$, under $M\leq 4$, is maximized at $M$ = 3 when $\alpha$ = 4.6021 and $M$ = 2 for the other values of $\alpha$. A similar argument can be applied to system (b), where the best rate performance is achieved with $M$ = 2 when $\rho$ = 5 dB and $M$ = 1 for the other cases. Figs. \ref{fig:fig5} and \ref{fig:fig5b} thus clearly confirm the conclusions inferred from Lemma \ref{lemma:theoDoF c-cell RBF-MMSE, RBF-MF/AS}. Note that the setting $M$ = 4 almost gives the worst rate performance in all cases.


For convenience, let $\mv{d}_{\text{RBF-Rx}}(\mv{\alpha}, \mv{m})$ = $\big[$$d_{\text{RBF-Rx},1}(\alpha_1, \mv{m})$, $\cdots$, $d_{\text{RBF-Rx},C}(\alpha_C, \mv{m})$$\big]^T$ be the DoF vector, with $d_{\text{RBF-Rx},c}(\alpha_c, \mv{m})$, $c=1, \dots, C$, obtained from Lemma \ref{lemma:theoDoF c-cell RBF-MMSE, RBF-MF/AS}. The characterization of the DoF region for the multi-cell RBF scheme with different receive spatial diversity techniques is then given in the following proposition.

\begin{proposition}\label{prop:DoF region}
Given $K_c=\Theta\left(\rho^{\alpha_c}\right), c=1,\ldots,C$, the achievable DoF region of a $C$-cell MIMO RBF system is given by
\begin{align}\label{eq:DoF region}
	\mathcal{D}_{\text{RBF-Rx}}(\mv{\alpha}) = \mv{conv} & \bigg\{\mv{d}_{\text{RBF-Rx}}(\mv{\alpha}, \mv{m}), M_c \in \{ 0, \cdots, N_T\}, c=1,\cdots,C\bigg\},
\end{align}
where \mv{conv} denotes the convex hull operation over all DoF vectors obtained with different values of $\mv{m}$ and ``Rx'' stands for MMSE, MF, or AS.
\end{proposition}

\begin{figure}[t]
    \centering
    \epsfxsize=0.65\linewidth
  	\includegraphics[width=8.5cm, height=8.5cm]{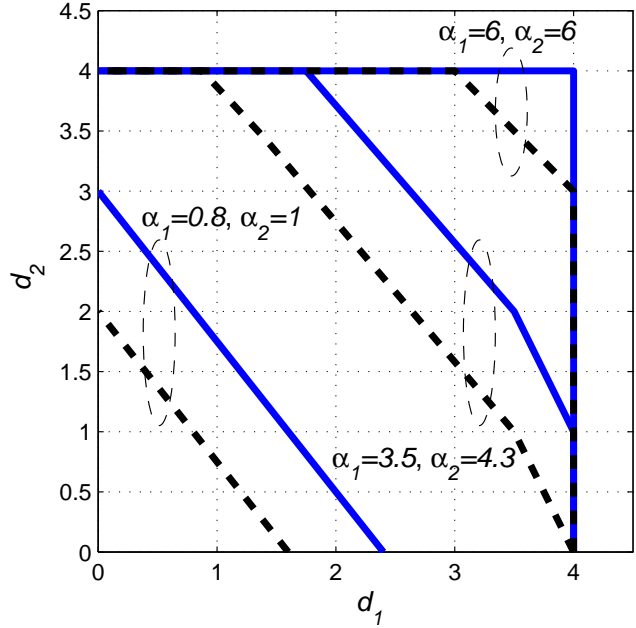}
  	\vspace{-0.1in}
  	\captionsetup{width=0.65\textwidth}
    \caption{DoF regions of two-cell MIMO RBF with different types of diversity receivers. The region boundaries for RBF-MMSE and RBF-MF/AS are denoted by solid and dashed lines, respectively.}\label{fig:DoF regions}
    \vspace{-0.2in}
\end{figure}

Fig. \ref{fig:DoF regions} shows the DoF region of a two-cell system employing either RBF-MMSE or RBF-MF/AS. We assume $N_T = 4$ and $N_R = 2$. It is observed that when $\alpha_1$ and $\alpha_2$ are small, the DoF region is more notably expanded by using MMSE receiver over MF/AS receiver. We conclude that receive spatial diversity is more beneficial when the numbers of users are relatively small. Note that to obtain $d_c$ DoF, the number of users in the $c$-th cell are at least in the order of  $\Theta \Big( \rho^{d_c \frac{\sum_{l=1}^{C}M_l-N_R}{M_c}} \Big)$ and $\Theta \Big( \rho^{d_c\frac{\sum_{l=1}^{C}M_l-1}{M_c}} \Big)$ with RBF-MMSE and RBF-MF/AS, respectively (cf. Lemmas \ref{lemma:theoDoF c-cell RBF-MMSE, RBF-MF/AS}). Thus, significantly less number of users per cell is required in RBF-MMSE as compared to RBF-MMF/AS for achieving the same DoF.

In practice, each cell can set different numbers of transmit beams at the BS. In general, the optimal DoF tradeoffs or the boundary DoF points are achieved when all the MSs apply the MMSE receiver and all the BSs cooperatively assign their numbers of transmit beams based on per-cell user densities and number of transmit/receive antennas. However, it is worth noting that there exists an underlying tradeoff between the achievable DoF and the receiver complexity, which determines the most desirable operating configuration of the system in consideration. 

\subsection{Optimality of Multi-Cell RBF}\label{sec:optimality}

It can be inferred from Lemma \ref{lemma:theoDoF c-cell RBF-MMSE, RBF-MF/AS} and observed from Fig. \ref{fig:DoF regions} that the DoF regions of RBF-MMSE and RBF-MF/AS both converge to the same region if the per-cell user densities are sufficiently large, in which all cells attain their maximum DoF $N_T$ by setting $M_c=N_T$, $\forall c$. The converged region is thus the ``interference-free'' DoF region as if there was no ICI such that each cell can be treated as an independent single-cell system. The above result implies that the multi-cell RBF is conceivably DoF-optimal given a sufficiently large number of users per cell, which is an extension of Proposition \ref{prop:optimality single-cell} to the multi-cell case. In this subsection, we rigorously develop this result. First, we present a (crude) DoF region upper bound for the $C$-cell MIMO downlink system with arbitrary transmission schemes. The proof follows directly from Proposition \ref{prop:single-cell DoF upperbound} and is thus omitted for brevity.

\begin{proposition}\label{prop:DoF region upperbound}
Given $K_c = \Theta(\rho^{\alpha_c})$, $c=1,\cdots,C$, an upper bound of the DoF region defined in (\ref{eq:def. DoF region}) for a $C$-cell MIMO downlink system is
\vspace{-0.1in}
\begin{align} \label{eq:DoFregion upper bound}
    \mathcal{D}_{\rm UB}(\mv{\alpha}) = \bigg\{ (d_1,d_2,\cdots\,d_C)\in\mathbb{R}_{+}^{C}: d_c\leq N_T, c=1,\cdots,C \bigg\}.
\end{align}
\end{proposition}

The DoF optimality of multi-cell RBF schemes is then obtained in the following proposition.

\begin{proposition}\label{prop:optimality of RBF-Rx}
Given $K_c = \Theta(\rho^{\alpha_c})$, $c=1,\cdots,C$, the multi-cell RBF schemes with different receive spatial diversity techniques achieve the ``interference-free'' DoF region upper bound of a $C$-cell MIMO downlink system, i.e., $\mathcal{D}_{\text{RBF-Rx}}(\mv{\alpha})$ $=$ $\mathcal{D}_{\rm UB}(\mv{\alpha})$, if 
\begin{itemize}
	\item RBF-MMSE: $\alpha_c \geq CN_T-N_R$, $\forall c\in \{ 1, \cdots, C\}$.
	\item RBF-MF/AS: $\alpha_c \geq CN_T-1$, $\forall c\in \{ 1, \cdots, C\}$.  
\end{itemize}
\end{proposition}

A direct consequence of Proposition \ref{prop:optimality of RBF-Rx} is thus  $\mathcal{D}_{\text{RBF-Rx}}(\mv{\alpha})$ $=$ $\mathcal{D}_{\text{MIMO}}(\mv{\alpha})$, i.e., each RBF scheme is indeed DoF-optimal when the numbers of users in all cells are sufficiently large. Due to the dominant multi-user diversity gain, RBF compensates the lack of full CSI at transmitters without any DoF loss. Furthermore, to achieve the interference-free DoF region, we infer from Proposition \ref{prop:optimality of RBF-Rx} that RBF-MMSE requires a much less number of users per cell, with a  difference of $N_R-1$ in the scaling order with respect to the SNR, as compared to RBF-MF or RBF-AS.

\section{Conclusion}\label{sec:conclusion}
This paper has studied the achievable sum-rate in multi-cell MIMO RBF systems for the regime of both high SNR and large number of users per cell. We propose three RBF schemes for spatial diversity receivers with multiple antennas, namely, RBF-MMSE, RBF-MF, and RBF-AS. The SINR distributions in the multi-cell RBF with different types of spatial receiver are obtained in closed-form at any given finite SNR. Based on these results, we characterize the DoF region achievable by different multi-cell MIMO RBF schemes under the assumption that the number of users per cell scales in a polynomial order with the SNR as the SNR goes to infinity. Our study reveals significant gains by using MMSE-based spatial receiver in the achievable sum-rate and DoF region in multi-cell RBF, which considerably differs from the existing result based on the conventional asymptotic analysis with fixed per-cell SNR. The results of this paper thus provide new insights on the optimal design of interference-limited multi-cell MIMO systems with only partial CSI at transmitters.

\appendices

\section{Preliminaries for the proof of Theorem \ref{theorem:general}}\label{appen:fact and lemmas}

In this appendix as well as the subsequent Appendix \ref{proof:general}, we use $\{ a_{i,j}\}_{i,j}$ to denote a matrix $\mv{A}$ having $a_{i,j}$ as the $(i,j)$-th component. The determinant and spectral norm of a symmetric matrix $\mv{A}$ are denoted by $|\mv{A}|$ and $||\mv{A}||_2$, respectively. $\mv{A} \succ \mv{0}$ means that $\mv{A}$ is a Hermitian and positive definite matrix. $\tensor*[_{p}]{F}{_q^{(n)}}(a_1,\ldots,a_p;b_1,\ldots,b_q;\mv{A})$ and $\tensor*[_{p}]{F}{_q^{(n)}}(a_1,\cdots,a_p;b_1,\cdots,b_q;\mv{A}, \mv{B})$ denote the hyper-geometric function of one and two matrix arguments, respectively \cite{James}. $\Gamma(m)$ and $\tilde{\Gamma}_m(n)$ are the gamma and complex multi-variate gamma function, respectively \cite{James}. $O(n)$ denotes the set of all orthogonal matrix with dimension $n$, and $[d\mv{U}]$ is the normalized Haar invariant probability measure on $O(n)$, normalized to make the total measure unity \cite{James}. $\text{etr}(\mv{X})$ is the short-hand notation for $\text{e}^{\mv{Tr}\left(\mv{X}\right)}$. $V(\mv{A})$ and $\Delta(\mv{A})$ denote the Vandermonde determinants of a diagonal matrix $\mv{A}=diag(a_1,\ldots,a_n)$, where $V(\mv{A})$ = $|\{a_i^{j-1}\}_{i,j=1,\ldots,n}|$ = $\prod_{1\leq i\leq j\leq n}(a_j - a_i)$ and $\Delta(\mv{A})$ = $|\{a_i^{n-j}\}_{i,j=1,\ldots,n}|$ = $\prod_{1\leq i\leq j\leq n}(a_i - a_j)$. Clearly, $\Delta(\mv{A})$ = $(-1)^{n(n-1)/2}V(\mv{A})$. Next, we present several lemmas that will be used to prove Theorem \ref{theorem:general} in Appendix \ref{proof:general}.

\begin{lemma}\label{lemma:sumA_i}
Suppose that $\psi_j\neq \psi_i$, $i\neq j \in \{1,\ldots,n\}$, and $A_i = 1/\prod_{j\neq i}^n (1-\psi_j/\psi_i)$. Then we have
\begin{align}\label{eq: A1 1}
	1 - \left( \prod_{i=1}^{n}\psi_i \right) \sum_{i=1}^n\frac{(-1)^{n+1}A_i}{\psi_i^n(1+\psi_i s)} = \frac{\prod_{i=1}^{n}\psi_i s^n}{\prod_{i=1}^n (1+\psi_i s)}.
\end{align}
\end{lemma}
\begin{IEEEproof}
Multiplying both sides of (\ref{eq: A1 1}) with $\prod_{i=1}^n(1+\psi_i s)$ and subtracting them, we obtain a polynomial in $s$ of degree $n$. However, since this polynomial has $n+1$ zeros: $s\in\{0,-1/\psi_1,\ldots,-1/\psi_n\}$, it should be equal to 0. This completes the proof of Lemma \ref{lemma:sumA_i}. 
\end{IEEEproof}

\begin{lemma}[\cite{James}](Splitting Property)\label{lemma:splitting}
Suppose that $\mv{A}$ $\in$ $\mathbb{C}^{n \times n}$, $\mv{A}\succ \mv{0}$,  and $\mv{B}$ $\in$ $\mathbb{C}^{n \times n}$ is a Hermitian matrix. Then we have
\begin{align}
	\int_{\mv{U}\in O(n)} \tensor*[_{p}]{F}{_q^{(n)}}(a_1,\cdots,a_p;b_1,\cdots,b_q;\mv{A}\mv{U}\mv{B}\mv{U}^H)[d\mv{U}] 
	= \tensor*[_{p}]{F}{_q^{(n)}}(a_1,\cdots,a_p;b_1,\cdots,b_q;\mv{A}, \mv{B}).	
\end{align}  
\end{lemma}

\begin{lemma}[\cite{James}](Reproductive Property)\label{lemma:reproductive}
Suppose that $\mv{A}$ $\in$ $\mathbb{C}^{n \times n}$, $\mv{A}\succ \mv{0}$, and $\mv{B}$, $\mv{C}$ $\in$ $\mathbb{C}^{n \times n}$ are Hermitian matrices. Then, for any complex number $a$ with the real part $\text{Re}(a)>n-1$, we have
\begin{align}
	& \int_{\mv{X}\succ\mv{0}} \text{etr}(-\mv{A}\mv{X})|\mv{X}|^{a-n}\tensor*[_{p}]{F}{_q^{(n)}}(a_1,\cdots,a_p;b_1,\cdots,b_q;\mv{X}\mv{B},\mv{C}) d\mv{X}\notag \\
	& \qquad \qquad = \tilde{\Gamma}_n(a)|\mv{A}|^{-a}\tensor*[_{p+1}]{F}{_q^{(n)}}(a_1,\cdots,a_p,a;b_1,\cdots,b_q;\mv{A}^{-1}\mv{B},\mv{C}).	
\end{align}  
\end{lemma}

\begin{lemma}[\cite{James}](Eigenvalue Transformation)\label{lemma:eigenvalue}
Suppose that $\mv{A}$ $\in$ $\mathbb{C}^{n \times n}$, $\mv{A}\succ \mv{0}$, is a Hermitian matrix with the joint distribution $f(\mv{A})$. The joint probability distribution function (PDF) of the eigenvalues $\lambda_n>\ldots>\lambda_1>0$ of $\mv{A}$ is 
\begin{align}
	g(\mv{\Lambda}) = \frac{\pi^{n(n-1)}}{\tilde{\Gamma}_n(n)} V^2(\mv{\Lambda}) \int_{\mv{U}\in O(n)} f(\mv{U}\mv{\Lambda}\mv{U}^H) [d\mv{U}],
\end{align}   
where $\mv{\Lambda} = diag(\lambda_1, \dots, \lambda_n)$ and $\mv{U}\mv{\Lambda}\mv{U}^H$ is the eigenvalue decomposition of the matrix $\mv{A}$.
\end{lemma}

\begin{lemma}[\cite{Khatri}](Quadratic Distribution)\label{lemma:quadratic}
Suppose that $\mv{X}\sim\mathcal{CN}(\mv{0}_{p\times n},\mv{\Sigma}\otimes\mv{\Omega})$, $n\geq p$, and $\mv{M}\succ \mv{0}$. The distribution of $\mv{V} = \mv{X}\mv{M}\mv{X}^H$ is
	\begin{align} 
		f_{\mv{V}}(\mv{V}) = \frac{1}{\tilde{\Gamma}_p(n) |\mv{\Sigma}|^n |\mv{M}\mv{\Omega}|^p }  |\mv{V}|^{n-p} \tensor*[_{0}]{F}{_0^{(n)}}(\mv{M}^{-1/2}\mv{\Omega}^{-1}\mv{M}^{-1/2}, -\mv{\Sigma}^{-1}\mv{V}).
	\end{align}
\end{lemma} 

\begin{lemma}\label{lemma:n=p}
Suppose that $\mv{\Psi}=diag(\psi_1,\ldots,\psi_p)$, $\psi_i > 0$, $\psi_j\neq \psi_i$, $i, j \in \{1,\ldots,n\}$, $i \neq j$, and
	\begin{align} 
		f_{S}(s) = \frac{s^{p-1}\tilde{\Gamma}_p(p+1)}{\tilde{\Gamma}_p(p)\tilde{\Gamma}(p)} |\mv{\Psi}| \tensor*[_{1}]{F}{_0^{(p)}}(p+1;\mv{\Psi},-s).
	\end{align}
The function $F_{S}(s)=\int_{0}^{s}f_{S}(x)dx$ is then in the form of
	\begin{align} 
		F_{S}(s) = \frac{\prod_{i=1}^{n}\psi_i s^n}{\prod_{i=1}^n (1+\psi_i s)}.
	\end{align}
\end{lemma} 
\begin{IEEEproof}
Note that $\tensor*[_{1}]{F}{_0^{(p)}}(p+1;\mv{\Psi},-s) = \displaystyle\lim_{\epsilon_1,\ldots,\epsilon_{p-2}\to 0}\tensor*[_{1}]{F}{_0^{(p)}}(p+1;\mv{\Psi},-\mv{S}_1)$, where $\mv{S}_1=diag(s,0,\epsilon_1,\ldots,\epsilon_{p-2})$. From \cite[(4.7)]{Gross}, we have
\begin{align}\label{eq:frac1}
	\tensor*[_{1}]{F}{_0^{(p)}}(p+1;\mv{\Psi},-\mv{S}_1) = \frac{|\tensor*[_{1}]{F}{_0}(2;-\psi_i s_j)|}{p!\Delta(\mv{\Psi}) \Delta(-\mv{S}_1)}
	= (-1)^{\frac{p(p-1)}{2}}\frac{|\tensor*[_{1}]{F}{_0}(2;-\psi_i s_j)|}{p!V(\mv{\Psi}) V(\mv{S}_1)},
\end{align}
where $|\tensor*[_{1}]{F}{_0}(2;-\psi_i s_j)|$ denotes the determinant of a matrix with $(i,j)$-th component being $\tensor*[_{1}]{F}{_0}(2;-\psi_i s_j)$ = $1/(1+\psi_i s_j)^2$, and $[s_1,\ldots,s_p] := [s,0,\epsilon_1,\ldots,\epsilon_{p-2}]$. Denote
\begin{align}
	|\tensor*[_{1}]{F}{_0}(2;-\psi_i s_j)| 
	= \begin{vmatrix} \frac{1}{(1+\psi_1 s)^2} & 1 & \frac{1}{(1+\psi_1 \epsilon_1)^2} & \cdots & \frac{1}{(1+\psi_1 \epsilon_{p-2})^2} \\
	\vdots & \vdots & \vdots & \vdots & \vdots \\
	\frac{1}{(1+\psi_p s)^2} & 1 & \frac{1}{(1+\psi_p \epsilon_1)^2} & \cdots & \frac{1}{(1+\psi_p \epsilon_{p-2})^2} \end{vmatrix}
	:= |\mv{c}_0,\mv{c}_1,\mv{g}(\epsilon_1),\ldots,\mv{g}(\epsilon_{p-2})|.
\end{align} 
Similarly, we have
\begin{align}
	V(\mv{S}_1) 
	= \begin{vmatrix} 1 & 1 & 1 & \cdots & 1 \\
	s & 0 & \epsilon_1 & \cdots & \epsilon_{p-2} \\
	\vdots & \vdots & \vdots & \vdots & \vdots \\
	s^{p-1} & 0 & \epsilon_1^{p-1} & \cdots & \epsilon_{p-2}^{p-1} \end{vmatrix}
	:= |\mv{c}_2,\mv{c}_3,\mv{h}(\epsilon_1),\ldots,\mv{h}(\epsilon_{p-2})|.
\end{align}
 
Using the \emph{L' Hospital rule} \cite[3.4.1]{Abramowitz}, we then have
\begin{align}\label{eq:frac2}
	\frac{|\tensor*[_{1}]{F}{_0}(2;-\psi_i s_j)|}{V(\mv{S}_1)} = 
	\frac{|\mv{c}_0,\mv{c}_1,\mv{g}(\epsilon_1),\ldots,\mv{g}(\epsilon_{p-2})|} {|\mv{c}_2,\mv{c}_3,\mv{h}(\epsilon_1),\ldots,\mv{h}(\epsilon_{p-2})|}
	= \frac{\left| \mv{c}_0,\mv{c}_1,\displaystyle\frac{d\mv{g}(x)}{dx}\Big|_{x=\epsilon_1}, \ldots, \displaystyle\frac{d^{p-2}\mv{g}(x)}{dx^{p-2}}\Big|_{x=\epsilon_{p-2}} \right|}  {\left| \mv{c}_2,\mv{c}_3,\displaystyle\frac{d\mv{h}(x)}{dx}\Big|_{x=\epsilon_1},\ldots,\displaystyle\frac{d^{p-2}\mv{h}(x)}{dx^{p-2}}\Big|_{x=\epsilon_{p-2}} \right|}.	
\end{align}

It is easy to see that
\begin{align}\label{eq:lim1}
	& \left| \mv{c}_0,\mv{c}_1,\displaystyle\frac{d\mv{g}(x)}{dx}\Big|_{x=\epsilon_1}, \ldots, \displaystyle\frac{d^{p-2}\mv{g}(x)}{dx^{p-2}}\Big|_{x=\epsilon_{p-2}} \right| 
	\xrightarrow{\epsilon_1,\ldots,\epsilon_{p-2}\to 0} \notag \\ 
	& \qquad \qquad \qquad \qquad \qquad (-1)^{\frac{(p-1)(p-2)}{2}}\left(\prod_{i=1}^{p}\Gamma(i)\right) 
	\begin{vmatrix} \frac{1}{(1+\psi_1 s)^2} & 1 & \psi_1 & \cdots & \psi_1^{p-2} \\
	\vdots & \vdots & \vdots & \vdots & \vdots \\
	\frac{1}{(1+\psi_p s)^2} & 1 & \psi_p & \cdots & \psi_p^{p-2} \end{vmatrix}, 	
\end{align}

\begin{align}\label{eq:lim2}
	\left| \mv{c}_2,\mv{c}_3, \displaystyle\frac{d\mv{h}(x)}{dx}\Big|_{x=\epsilon_1}, \ldots, \displaystyle\frac{d^{p-2}\mv{h}(x)}{dx^{p-2}}\Big|_{x=\epsilon_{p-2}} \right| \xrightarrow{\epsilon_1,\ldots,\epsilon_{p-2}\to 0} 
	\begin{vmatrix} 1 & 1 & 0 & \cdots & 0 \\
	s & 0 & 1 & \cdots & 0 \\
	\vdots & \vdots & \vdots & \vdots & \vdots \\
	s^{p-2} & 0 & 0 & \cdots & (p-2)! \\
	s^{p-1} & 0 & 0 & \cdots & 0  \end{vmatrix} \notag \\ 
	= (-1)^{p-1}\left(\prod_{i=1}^{p-1}\Gamma(i)\right)s^{p-1}.
\end{align}  

Combining (\ref{eq:frac1}), (\ref{eq:frac2}), (\ref{eq:lim1}), and (\ref{eq:lim2}), $f_S(s)$ can be expressed as 
\begin{align}
	f_{S}(s) = \frac{1}{V(\mv{\Psi})}
	\begin{vmatrix} \frac{\psi_1}{(1+\psi_1 s)^2} & \psi_1 & \cdots & \psi_1^{p-1} \\
	\vdots & \vdots & \vdots & \vdots  \\
	\frac{\psi_p}{(1+\psi_p s)^2} & \psi_p & \cdots & \psi_p^{p-1} \end{vmatrix}.
\end{align}

Hence,
\begin{align}
	F_{S}(s) = \int_{0}^{s}f_S(x)dx = \frac{1}{V(\mv{\Psi})}
	\begin{vmatrix} 1- \frac{1}{1+\psi_1 s} & \psi_1 & \cdots & \psi_1^{p-1} \\
	\vdots & \vdots & \vdots & \vdots  \\
	1- \frac{1}{1+\psi_p s} & \psi_p & \cdots & \psi_p^{p-1} \end{vmatrix}
	= 1 - \frac{1}{V(\mv{\Psi})}
	\begin{vmatrix} \frac{1}{1+\psi_1 s} & \psi_1 & \cdots & \psi_1^{p-1} \\
	\vdots & \vdots & \vdots & \vdots  \\
	\frac{1}{1+\psi_p s} & \psi_p & \cdots & \psi_p^{p-1} \end{vmatrix}.
\end{align}

Using the \emph{Laplace's cofactor expansion} \cite[14.15]{Gradshteyn}, we obtain
\begin{align}
	F_{S}(s) 
	& = 1 - \sum_{i=1}^p\frac{(-1)^{i+1}\prod_{j\neq i}\psi_j}{1+\psi_i s}\frac{V(\mv{\Psi_{-i}})}{V(\mv{\Psi})}
	= 1 - \sum_{i=1}^p\frac{(-1)^{i+1}\prod_{j\neq i}\psi_j}{1+\psi_i s}\frac{1}{\prod_{j<i}(\psi_i-\psi_j)\prod_{j>i}(\psi_j-\psi_i)} \notag \\
	& = 1 - \left( \prod_{i=1}^{p}\psi_i \right) \sum_{i=1}^p\frac{(-1)^{p+1}A_i}{\psi_i^p(1+\psi_i s)}
	= \frac{\prod_{i=1}^{p}\psi_i s^p}{\prod_{i=1}^p (1+\psi_i s)}, 
\end{align}
where $A_i = 1/\prod_{j\neq i}^p (1-\psi_j/\psi_i)$, $\mv{\Psi}_{-i}=diag(\psi_1,\ldots,\psi_{i-1},\psi_{i+1},\ldots,\psi_p)$, and the last equality is obtained from Lemma \ref{lemma:sumA_i}. This completes the proof of Lemma \ref{lemma:n=p}.
\end{IEEEproof}

\section{Proof of Theorem \ref{theorem:general}}\label{proof:general}
Denoting $\mv{V} = \mv{X}\mv{\Psi}\mv{X}^H$, the PDF of $\mv{V}$ can then be obtained from Lemma \ref{lemma:quadratic} as 
\begin{align}\label{eq:fv}
	f_{\mv{V}}(\mv{V}) = \frac{|\mv{V}|^{n-p}}  {\tilde{\Gamma}_p(n)|\mv{\Psi}|^p}  \tensor*[_{0}]{F}{_0^{(n)}}(\mv{\Psi}^{-1},-\mv{V}) d\mv{V}.  
\end{align}

From Lemma \ref{lemma:quadratic}, given $\mv{V}$, the conditional PDF of the random variable $S:=\mv{h}^H\mv{V}^{-1}\mv{h}$ can be expressed as
\begin{align}\label{eq:fs_v}
	f_{S|\mv{V}}(s|\mv{V}) = \frac{|\mv{V}|s^{p-1}}{\Gamma(p)} \tensor*[_{0}]{F}{_0^{(p)}}(\mv{V},-s).
\end{align}

From (\ref{eq:fv}) and (\ref{eq:fs_v}), we have
\begin{align}\label{eq:general1}
	f_S(s) = \int_{\mv{V}\succ \mv{0}} f_{S|\mv{V}}(s|\mv{V}) f_{\mv{V}}(\mv{V}) d\mv{V} 
	= \frac{s^{p-1}}{\Gamma(p)\tilde{\Gamma}_p(n)|\mv{\Psi}|^p} \int_{\mv{V}\succ \mv{0}} |\mv{V}|^{n+1-p} \tensor*[_{0}]{F}{_0^{(p)}}(\mv{V},-s) \tensor*[_{0}]{F}{_0^{(n)}}(\mv{\Psi}^{-1},-\mv{V}) d\mv{V}.  
\end{align}

Next, we prove Theorem \ref{theorem:general} by induction. We first prove that Theorem \ref{theorem:general} is true for $n=p$, and then given that it is true for $n=m-1\geq p$, we show that it is also true for $n=m$. Note that for convenience, we assume $\psi_j\neq \psi_i$, $i\neq j \in \{1,\ldots,n\}$. 

\subsection{The Case of $n=p$}
From (\ref{eq:general1}) with $n=p$, we have
\begin{align} 
	f_S(s) 
	&= \frac{s^{p-1}}{\Gamma(p)\tilde{\Gamma}_p(p)|\mv{\Psi}|^p} \int_{\mv{V}\succ \mv{0}} |\mv{V}| \tensor*[_{0}]{F}{_0^{(p)}}(\mv{V},-s) \tensor*[_{0}]{F}{_0^{(p)}}(\mv{\Psi}^{-1},-\mv{V}) d\mv{V}  \notag \\
	&= \frac{s^{p-1}}{\Gamma(p)\tilde{\Gamma}_p(p)|\mv{\Psi}|^p} \int_{\mv{V}\succ \mv{0}} \int_{\mv{U}\in O(p)} |\mv{V}| \text{etr}(-\mv{U}\mv{\Psi}^{-1}\mv{U}^H\mv{V}) \tensor*[_{0}]{F}{_0^{(p)}}(\mv{V},-s) [d\mv{U}] d\mv{V}  \label{eq:up} \\
	& = \frac{s^{p-1}\tilde{\Gamma}_p(p+1)}{\Gamma(p)\tilde{\Gamma}_p(p)|\mv{\Psi}|^p}  \int_{\mv{U}\in O(p)} |\mv{U}\mv{\Psi}^{-1}\mv{U}^H|^{-(n+1)} \tensor*[_{1}]{F}{_0^{(p)}}(p+1; \left(\mv{U}\mv{\Psi}^{-1}\mv{U}^H\right)^{-1},-s) [d\mv{U}]
	\end{align}
	\begin{align}
	&= \frac{s^{p-1}\tilde{\Gamma}_p(p+1)}{\Gamma(p)\tilde{\Gamma}_p(p)} |\mv{\Psi}| \tensor*[_{1}]{F}{_0^{(p)}}(p+1; \mv{\Psi},-s), \label{eq:down} 
\end{align}
where (\ref{eq:up}) follows from  Lemma \ref{lemma:reproductive}. Combining (\ref{eq:down}) and Lemma \ref{lemma:n=p}, it follows that Theorem \ref{theorem:general} is true for $n=p$.  

\subsection{The Case of $n > p$} 
Suppose that Theorem \ref{theorem:general} is true for $n=m-1>p$. We will show in the following that it is also true for $n=m$. Applying Lemma \ref{lemma:eigenvalue} to (\ref{eq:general1}) with any pair of $n$ and $m$ with $n=m-1>p$, we have
\begin{align}
	f_S(s) = \frac{s^{p-1}}{\Gamma(p)\tilde{\Gamma}_p(m)|\mv{\Psi}|^p} \frac{\pi^{p(p-1)}}{\tilde{\Gamma}_p(p)} \int_{\infty>\lambda_p>\ldots>\lambda_1>0} V^2(\mv{\Lambda}) |\mv{\Lambda}|^{m+1-p} \tensor*[_{0}]{F}{_0^{(p)}}(\mv{\Lambda},-s) \tensor*[_{0}]{F}{_0^{(m)}}(\mv{\Psi}^{-1},-\mv{\Lambda}) d\mv{\Lambda},
\end{align}   
where $\mv{\Lambda}=diag(\lambda_1,\ldots,\lambda_p)$. We then find an alternative form for $\tensor*[_{0}]{F}{_0^{(m)}}(\mv{\Psi}^{-1},-\mv{\Lambda})$ by using \cite[(4.6)]{Gross} as follows:
\begin{align}
	\tensor*[_{0}]{F}{_0^{(m)}}(\mv{\Psi}^{-1},-\mv{\Lambda}) = \lim_{\delta_1,\ldots,\delta_{m-p-1}\to 0} \tensor*[_{0}]{F}{_0^{(m)}}(\mv{\Psi}^{-1},-\mv{\Lambda_1}) 
	= (-1)^{\frac{m(m-1)}{2}} \prod_{k=1}^m \Gamma(k) \lim_{\delta_1,\ldots,\delta_{m-p-1}\to 0}  \frac{|\tensor*[_{0}]{F}{_0}(-\psi_i^{-1}\lambda_j)|}{V(\mv{\Psi}^{-1})V(\mv{\Lambda}_1)},
\end{align}
where $\mv{\Lambda}_1 = diag(\lambda_1,\ldots,\lambda_m) := diag(\lambda_1,\ldots,\lambda_p,0,\delta_1,\ldots,\delta_{m-p-1})$. Denote 
\begin{align}
	\left| \tensor*[_{0}]{F}{_0}(-\frac{\lambda_j}{\psi_i}) \right|
	& = \begin{vmatrix}
	\text{exp}(-\lambda_1/\psi_1) & \cdots & \text{exp}(-\lambda_p/\psi_1) & 1 & \text{exp}(-\delta_1/\psi_1) & \cdots & \text{exp}(-\delta_{m-p-1}/\psi_1) \\
	\text{exp}(-\lambda_1/\psi_2) & \cdots & \text{exp}(-\lambda_p/\psi_2) & 1 & \text{exp}(-\delta_1/\psi_2) & \cdots & \text{exp}(-\delta_{m-p-1}/\psi_2) \\
	\vdots & \vdots & \vdots & \vdots & \vdots & \vdots & \vdots \\
	\text{exp}(-\lambda_1/\psi_m) & \cdots & \text{exp}(-\lambda_p/\psi_m) & 1 & \text{exp}(-\delta_1/\psi_m) & \cdots & \text{exp}(-\delta_{m-p-1}/\psi_m)
	\end{vmatrix} \notag \\ 
	& := |\mv{d}_0,\ldots,\mv{d}_p, \mv{g}_1(\delta_1), \ldots, \mv{g}_1(\delta_{m-p-1})|, 	
\end{align}

and
\begin{align}
	V(\mv{\Lambda}_1)
	& = \begin{vmatrix}
	1 & \cdots & 1 & 1 & 1 & \cdots & 1 \\
	\lambda_1 & \cdots & \lambda_p & 0 & \delta_1 & \cdots & \delta_{m-p-1} \\
	\vdots & \vdots & \vdots & \vdots & \vdots & \vdots & \vdots \\
	\lambda_1^{m-1} & \cdots & \lambda_p^{m-1} & 0 & \delta_1^{m-1} & \cdots & \delta_{m-p-1}^{m-1} \\ 
	\end{vmatrix} 
	& := |\mv{d}_{p+1},\ldots,\mv{d}_{2p+1}, \mv{h}_1(\delta_1), \ldots, \mv{h}_1(\delta_{m-p-1})|.	
\end{align}

Using the \emph{L' Hospital rule} \cite[3.4.1]{Abramowitz}, we have
\begin{align}\label{eq:x1}
	\frac{|\tensor*[_{0}]{F}{_0}(-\frac{\lambda_j}{\psi_i})|}{V(\mv{\Lambda}_1)} 
	&= \frac{|\mv{d}_0,\ldots,\mv{d}_p, \mv{g}_1(\delta_1), \ldots, \mv{g}_1(\delta_{m-p-1})|} 
	{|\mv{d}_{p+1},\ldots,\mv{d}_{2p+1}, \mv{h}_1(\delta_1), \ldots, \mv{h}_1(\delta_{m-p-1})|} \notag \\
	&= \frac{\left| \mv{d}_0,\ldots,\mv{d}_p, \displaystyle\frac{d\mv{g}_1(x)}{dx}\Big|_{x=\delta_1}, \ldots, \displaystyle\frac{d^{m-p-1}\mv{g}_1(x)}{dx^{m-p-1}}\Big|_{x=\delta_{m-p-1}} \right|}
	{\left|  \mv{d}_{p+1},\ldots,\mv{d}_{2p+1}, \displaystyle\frac{d\mv{h}_1(x)}{dx}\Big|_{x=\delta_1}, \ldots, \displaystyle\frac{d^{m-p-1}\mv{h}_1(x)}{dx^{m-p-1}}\Big|_{x=\delta_{m-p-1}} \right|}.	
\end{align}

It is easy to see in (\ref{eq:x1}) that
\begin{align}
	& \left| \mv{d}_{p+1},\ldots,\mv{d}_{2p+1}, \displaystyle\frac{d\mv{h}_1(x)}{dx}\Big|_{x=\delta_1}, \ldots, \displaystyle\frac{d^{m-p-1}\mv{h}_1(x)}{dx^{m-p-1}}\Big|_{x=\delta_{m-p-1}} \right| \notag \\
	& \xrightarrow{\delta_1,\ldots,\delta_{m-p-1}\to 0} 
	\begin{vmatrix}
	1 & \cdots & 1 & 1 & 0 & \cdots & 0 & 0\\
	\lambda_1 & \cdots & \lambda_p & 0 & 1 & \cdots & 0 & 0\\
	\vdots & \vdots & \vdots & \vdots & \vdots & \vdots & \vdots & \vdots \\
	\lambda_1^{m-p-2} & \cdots & \lambda_p^{m-p-2} & 0 & 0 & \cdots & (m-p-2)! & 0 \\
	\lambda_1^{m-p-1} & \cdots & \lambda_p^{m-p-1} & 0 & 0 & \cdots & 0 & (m-p-1)! \\
	\lambda_1^{m-p} & \cdots & \lambda_p^{m-p} & 0 & 0 & \cdots & 0 & 0 \\
	\vdots & \vdots & \vdots & \vdots & \vdots & \vdots & \vdots & \vdots \\
	\lambda_1^{m-1} & \cdots & \lambda_p^{m-1} & 0 & 0 & \cdots & 0 & 0
	\end{vmatrix} \notag \\
	& = (-1)^{(p+2)(m-p)}\prod_{k=1}^{m-p}\Gamma(k) |\Lambda|^{m-p} V(\mv{\Lambda}), 	
\end{align}

and
\begin{align}
	& \left| \mv{d}_0,\ldots,\mv{d}_p, \displaystyle\frac{d\mv{g}_1(x)}{dx}\Big|_{x=\delta_1}, \ldots, \displaystyle\frac{d^{m-p-1}\mv{g}_1(x)}{dx^{m-p-1}}\Big|_{x=\delta_{m-p-1}} \right| \notag \\
	& \xrightarrow{\delta_1,\ldots,\delta_{m-p-1}\to 0} (-1)^{\frac{(m-p)(m-p-1)}{2}} 
	\begin{vmatrix} 
	\text{exp}(-\lambda_1/\psi_1) & \cdots & \text{exp}(-\lambda_p/\psi_1) & 1 & 1/\psi_1 & \cdots & 1/\psi_1^{m-p-1} \\
	\vdots & \vdots & \vdots & \vdots & \vdots \\
	\text{exp}(-\lambda_1/\psi_m) & \cdots & \text{exp}(-\lambda_p/\psi_m) & 1 & 1/\psi_m & \cdots & 1/\psi_m^{m-p-1}
	\end{vmatrix} \notag \\ 
	& \qquad \qquad \qquad ~ := (-1)^{\frac{(m-p)(m-p-1)}{2}} T(p,m,\lambda_1,\ldots,\lambda_p,\psi_1,\ldots,\psi_m),	
\end{align}
where the function $T(p,m,\lambda_1,\ldots,\lambda_p,\psi_1,\ldots,\psi_m)$ is defined for the sake of brevity. 

The PDF $f_S(s)$ can thus be expressed as
\begin{align}
	f_S(s) 
	= \frac{(-1)^{\frac{p(p-1)}{2}}s^{p-1}}{\Gamma(p)\prod_{k=1}^p\Gamma(k)V(\mv{\Psi}^{-1})|\mv{\Psi}|^p} 
	\int_{\infty>\lambda_p>\cdots>\lambda_1>0} V(\mv{\Lambda})|\mv{\Lambda}| \tensor*[_{0}]{F}{_0^{(p)}}(\mv{\Lambda},-s) T(p,m,\lambda_1,\ldots,\lambda_p,\psi_1,\ldots,\psi_m).
\end{align}

Now using the Laplace's cofactor expansion \cite[14.15]{Gradshteyn}, we can rewrite $T(p,m,\lambda_1,\ldots,\lambda_p,\psi_1,\ldots,\psi_m)$ as
\begin{align}  
	T(p,m,\lambda_1,\ldots,\lambda_p,\psi_1,\ldots,\psi_m) 
	= \displaystyle\sum_{i=1}^m \frac{(-1)^{m+i}}{\psi_i^{m-p-1}} T(p,m-1,\lambda_1,\ldots,\lambda_p,\psi_1,\ldots,\psi_{i-1},\psi_{i+1},\ldots,\psi_m).
\end{align}

Therefore, we have
\begin{align}
	F_S(s) 
	& = \displaystyle \sum_{i=1}^m \frac{(-1)^{m+i}}{\psi_i^{m-p-1}} \frac{V(\mv{\Psi}_{-i}^{-1})|\mv{\Psi}_{-i}|^p}{V(\mv{\Psi}^{-1})|\mv{\Psi}|^p} \frac{(-1)^{\frac{p(p-1)}{2}}s^{p-1}}{\Gamma(p)\prod_{k=1}^p\Gamma(k)V(\mv{\Psi}_{-i}^{-1})|\mv{\Psi}_{-i}|^p} \times \notag \\ 
	& \times \int_0^s\int_{\infty>\lambda_p>\cdots>\lambda_1>0} V(\mv{\Lambda})|\mv{\Lambda}| \tensor*[_{0}]{F}{_0^{(p)}}(\mv{\Lambda},-x) T(p,m-1,\lambda_1,\ldots,\lambda_p,\psi_1,\ldots,\psi_{i-1},\psi_{i+1},\ldots,\psi_m) \label{eq:note0} \\ 	
	& = \displaystyle \sum_{i=1}^m \frac{(-1)^{m+i}}{\psi_i^{m-p-1}} \frac{V(\mv{\Psi}_{-i}^{-1})|\mv{\Psi}_{-i}|^p}{V(\mv{\Psi}^{-1})|\mv{\Psi}|^p}
	\left[ \frac{\sum_{k=p}^{m-1}\beta_{k,-i}s^k}{\prod_{j=1,j\neq i}^{m}(1+\psi_j s)}\right]  
	= \prod_{k=1}^m\psi_k \displaystyle \sum_{i=1}^m \frac{(-1)^{m+1}A_i}{\psi_i^m} \left[ \frac{\sum_{k=p}^{m-1}\beta_{k,-i}s^k}{\prod_{j=1,j\neq i}^{m}(1+\psi_j s)}\right] \notag \\
	& = \frac{1}{\prod_{j=1}^{m}(1+\psi_j s)} \prod_{k=1}^m\psi_k \displaystyle \sum_{i=1}^m \frac{(-1)^{m+1}A_i}{\psi_i^m} \left[ \left( \sum_{k=p}^{m-1}\beta_{k,-i}s^k \right)(1+\psi_i s)\right] \label{eq:note1}, 
\end{align}
where (\ref{eq:note0}) follows due to the inductive assumption that Theorem \ref{theorem:general} is true for $n=m-1\geq p$.

Furthermore, from Lemma \ref{lemma:sumA_i}, we have 
\begin{align}\label{eq:note2}
	\frac{\prod_{i=1}^m(1+\psi_i s)}{\prod_{i=1}^m(1+\psi_i s)} 
	& = 1 = \prod_{k=1}^m\psi_k \displaystyle \sum_{i=1}^m \frac{(-1)^{m+1}A_i}{\psi_i^m}
	= \prod_{k=1}^m\psi_k \displaystyle \sum_{i=1}^m \frac{(-1)^{m+1}A_i}{\psi_i^m} \frac{\prod_{j=1,j\neq i}^m(1+\psi_j s)}{\prod_{j=1,j\neq i}^m(1+\psi_j s)} \notag \\
	& = \frac{1}{\prod_{j=1}^{m}(1+\psi_j s)} \prod_{k=1}^m\psi_k \displaystyle \sum_{i=1}^m \frac{(-1)^{m+1}A_i}{\psi_i^m} \left[ \left( \sum_{k=0}^{m-1}\beta_{k,-i}s^k \right)(1+\psi_i s)\right].
\end{align} 

By comparing (\ref{eq:note1}) and (\ref{eq:note2}), it follows that
\begin{align}
	F_S(s) = \frac{\sum_{k=p}^{m}\beta_k s^k}{\prod_{i=1}^m (1+\psi_i s)}.	
\end{align}  

Therefore, given that Theorem \ref{theorem:general} is true for $n=m-1\geq p$, it is also true for $n=m$. By combining the results in the above two cases, the proof of Theorem \ref{theorem:general} is thus completed.

\section{Proof of Corollary \ref{corollary:SINR RBF-MMSE CDF}}\label{proof:SINR RBF-MMSE CDF}
The interference-plus-noise covariance matrix $\mv{W}_k^{(c)}$ given in (\ref{eq:INCM}) can be written as
	\begin{align} 
		\mv{W}_k^{(c)} = \lim_{N\to\infty} \left(\frac{P_T}{M_c}\mv{\tilde{H}}_{k,-m}^{(c,c)}\Big(\mv{\tilde{H}}_{k,-m}^{(c,c)}\Big)^H + \displaystyle\sum_{l=1,l\neq c}^{C}\frac{P_T\gamma_{l,c}}{M_l}\mv{\tilde{H}}_{k}^{(l,c)}\Big(\mv{\tilde{H}}_{k}^{(l,c)}\Big)^H + \frac{\sigma^2}{N} \mv{H}_N \mv{H}_N^H \right), 
	\end{align}
where $\mv{H}_N \in \mathbb{C}^{N_R\times N}$ consists of i.i.d. random variables each distributed as $\sim\mathcal{CN}(0,1)$. To find the PDF of $\text{SINR}_{k,m}^{(\text{MMSE},c)}$ in (\ref{eq:SINR RBF-MMSE}), we apply Theorem \ref{theorem:general} with $\mv{h}:=\mv{\tilde{h}}_{k,m}^{(c,c)}$, 
\begin{align}
\mv{X} := \left[ \mv{\tilde{H}}_{k,-m}^{(c,c)}, \mv{\tilde{H}}_{k}^{(1,c)}, \cdots, \mv{\tilde{H}}_{k}^{(l,c)}, \cdots, \mv{\tilde{H}}_{k}^{(C,c)}, \mv{H}_N\right],
\end{align}

and 
\begin{align}
\mv{\Psi} := diag \left( \underbrace{1, \cdots, 1}_{M_c-1}, \cdots, \underbrace{ \frac{\mu_{l,c}}{\eta_c}, \cdots, \frac{\mu_{l,c}}{\eta_c} }_{M_l}, \cdots,  \underbrace{ \frac{\mu_{C,c}}{\eta_c}, \cdots, \frac{\mu_{C,c}}{\eta_c} }_{M_C}, \underbrace{\frac{1}{N\eta_c}, \cdots, \frac{1}{N\eta_c}}_{N}  \right).
\end{align}

The PDF of $S:=\text{SINR}_{k,m}^{(\text{MMSE},c)}$ can thus be expressed as
\begin{align}\label{eq:y1}
F_S(s) = 1 - \lim_{N\to\infty} \frac{ \left( \sum_{i=0}^{N_R - 1}\theta_i s^i \right) } { \left( 1+ \frac{s}{N\eta_c} \right)^N  (1+ s)^{M_c-1} \prod_{l=1, l\neq c}^{\sum M_c - 1} (1+ \frac{\mu_{l,c}}{\eta_c} s)^{M_l}},
\end{align}
where $\theta_i$ is the coefficient of $s^i$ in the polynomial expansion of $\left(1+ \frac{s}{N\eta_c}\right)^{N}(1+ s)^{M_c-1} \prod_{l=1, l\neq c}^{\sum M_c - 1} (1+ \frac{\mu_{l,c}}{\eta_c} s)^{M_l}$.

Next, by letting $N\to\infty$, in the denominator in (\ref{eq:y1}), the term $(1+ \frac{s}{N\eta_c})^{N}$ converges to $e^{s/\eta_c}$, while the nominator converges to $\sum_{i=0}^{N_R - 1}\zeta_i s^i$, where $\zeta_i$'s are defined in Corollary \ref{corollary:SINR RBF-MMSE CDF}. We thus obtain (\ref{eq:SINR RBF-MMSE CDF}). This completes the proof of Corollary \ref{corollary:SINR RBF-MMSE CDF}.

\section{Proof of Theorem \ref{theorem:SINR RBF-MF CDF}}\label{proof:SINR RBF-MF CDF}
We first note that
\begin{align}
	f_{S}^{(c)}(s) 
	& = \displaystyle\int_{0}^{\infty}{f_{S|V}(s|v)f_{V}(v)dv} \notag \\
	& = \int_{-\infty}^{\infty} \int_{0}^{\infty} \frac{s^{N_R-1}}{2\pi\Gamma(N_R)} \frac{ (v+\frac{1}{\eta_c})^{N_R} e^{-(v + \frac{1}{\eta_c})s} e^{-j\omega v}} {\left(1-j\omega\right)^{M_c-1}\displaystyle\prod_{l=1,l\neq c}^{C}
    {\left(1-j\frac{\mu_{l,c}}{\eta_c}\omega\right)^{M_l}}} dv d\omega,
\end{align}
where $j=\sqrt{-1}$.

Therefore, 
\begin{align}\label{eq: eq1 SINR RBF-MF CDF}
	F_S(s) = \int_{-\infty}^{\infty} \int_{0}^{\infty} \int_{0}^{s} \frac{ (v + \frac{1}{\eta_c})^{N_R} e^{-j\omega v} x^{N_R-1} e^{-(v + \frac{1}{\eta_c})x} } 
	{   2\pi \Gamma(N_R) \left(1-j\omega\right)^{M_c-1}\displaystyle\prod_{l=1,l\neq c}^{C}
    {\left(1-j\frac{\mu_{l,c}}{\eta_c}\omega\right)^{M_l}}   } dx dv d\omega. 
\end{align}


Now by using \cite[(3.351.1)]{Gradshteyn}, we can write (\ref{eq: eq1 SINR RBF-MF CDF}) as 
\begin{align}\label{eq: eq2 SINR RBF-MF CDF}
	F_S(s) & = 1 - \displaystyle\sum_{k=0}^{N_R-1} \frac{e^{-s/\eta_c}s^k}{2\pi k!} \int_{-\infty}^{\infty} \int_{0}^{\infty}
	\frac{ (v + \frac{1}{\eta_c})^{k} e^{-(s+j\omega)} }   { \left(1-j\omega\right)^{M_c-1}\displaystyle\prod_{l=1,l\neq c}^{C}
    {\left(1-j\frac{\mu_{l,c}}{\eta_c}\omega\right)^{M_l}} } dv d\omega
\\ \label{eq: eq3 SINR RBF-MF CDF}
   & = 1 - \displaystyle\sum_{k=0}^{N_R-1}\sum_{m=0}^{k} \frac{e^{-s/\eta_c}s^k}{(k-m)!\eta_c^{k-m}} 
  \underbrace{  \frac{1}{2\pi}   \int_{-\infty}^{\infty} \frac{d\omega} { (s+j\omega)^{m+1} \left(1-j\omega\right)^{M_c-1} \displaystyle\prod_{l=1,l\neq c}^{C} {\left(1-j\frac{\mu_{l,c}}{\eta_c}\omega\right)^{M_l}} }  }_{T_m(s)}, 
\end{align}
where we have used the binomial expansion and the result in \cite[(3.351.3)]{Gradshteyn} to obtain (\ref{eq: eq2 SINR RBF-MF CDF}). From \cite[(30)-(34)]{Hieu01}, we see that $T_0(s)$ can be expressed as in (\ref{eq:T_0}). It is also easy to show that $T_m(s) = \frac{(-1)^m}{m!} \frac{d^mT_0(s)}{ds^m}$. Combining this result, (\ref{eq:T_0}), and (\ref{eq: eq3 SINR RBF-MF CDF}), we obtain (\ref{eq:SINR RBF-MF CDF}). This completes the proof of Theorem \ref{theorem:SINR RBF-MF CDF}.

\section{Proof of Lemma \ref{lemma:theoDoF single-cell RBF-MMSE, RBF-MF/AS}}\label{proof:theoDoF single-cell RBF-MMSE, RBF-MF/AS}

\subsection{RBF-MMSE}
We first investigate the DoF with RBF-MMSE. Consider the following two cases.

\subsubsection{Case 1, $N_R\leq M - 1$}\label{subproof:theoDoF single-cell case 1}
Denote $R_{k,m}^{(\text{MMSE})} := \log_2\left(1+\text{SINR}_{k,m}^{(\text{MMSE})}\right)$. We first show that
\vspace{-0.15in}\begin{align}\label{eq:proof01_1a}
	\text{Pr}\left\{\frac{\alpha}{M - N_R}\log_2\eta +\log_2\log\eta \geq 
	\max_{k \in \{ 1, \cdots, K\}}R_{k,1}^{(\text{MMSE})} \right.
	\left. \geq \frac{\alpha}{M - N_R}\log_2\eta -\log_2\log\eta \right\} 
	\xrightarrow{\eta\to\infty}1,~ \text{if}~ 0< \alpha \leq M - N_R,
\end{align}
\begin{align}\label{eq:proof01_1b}
	\text{Pr}\bigg\{ \log_2\eta +\log_2\log\eta +\log_2\alpha\geq 
	\max_{k \in \{ 1, \cdots, K\}}R_{k,1}^{(\text{MMSE})} 
	\geq \log_2\eta +\log_2\log\eta +\log_2\beta_1 \bigg\} 
	\xrightarrow{\eta\to\infty}1,~ \text{if}~ \alpha > M - N_R,
\end{align}
where $\beta_1=\frac{\alpha-M+N_R}{2}$; hence, $\alpha>\beta_1>0$ when $\alpha$ $>$ $M - N_R$. From Corollary \ref{corollary:SINR RBF-MMSE CDF}, the CDF of the single-cell RBF-MMSE $S:=\text{SINR}_{k,1}^{(\text{MMSE})}$ is
\begin{align}
F_S(s) = 1 - e^{-s/\eta} \frac{\sum_{i=1}^{N_R-1}\frac{(M-1)!}{i!(M-1-i)!}s^i}{(1+s)^{M-1}} = 1 - e^{-s/\eta}\left( \Theta\left(\frac{1}{(s+1)^{M-N_R}} \right) + O\left( \frac{1}{(s+1)^{M-N_R+1}} \right)  \right),
\end{align} 
as $s$ and/or $\eta$ $\to\infty$. Therefore, the CDF of  $Y_k:=R_{k,1}^{(\text{MMSE})}$ has the following asymptotic form 
\begin{align}
        F_{Y_k}(y) = 1 - e^{-\left(2^y-1\right)/\eta} \left( \Theta\left(\frac{1}{2^{(M-N_R)y}} \right) + O\left( \frac{1}{2^{(M-N_R+1)y}} \right)  \right), \notag
\end{align}
as $y$ and/or $\eta$ $\to\infty$. In (\ref{eq:proof01_1a}), the upper-bound 
probability can thus be given as
\begin{align}\label{eq:eq 1}
    & \text{Pr}\left\{\frac{\alpha}{M-N_R}\log_2\eta +\log_2\log\eta\geq \max_{k \in \{ 1, \cdots, K\}}Y_k \right\}
    = \bigg[ F_{Y_k}\left( \frac{\alpha}{M-N_R}\log_2\eta +\log_2\log\eta \right) \bigg]^K \notag \\
    & = \vast(  1 - \exp{\left( -\eta^{\frac{\alpha}{M-N_R}-1}\log\eta +\frac{1}{\eta} \right)}  \left( \Theta\left(\frac{1}{\eta^{\alpha}\left( \log\eta \right)^{M-N_R} } \right) + O\left( \frac{1}{\left( \eta^{\frac{\alpha}{M-N_R}} \log\eta \right)^{M-N_R+1}} \right) \right) \vast)^K,
\end{align}
as $\eta\to\infty$. Note that when $x$ is small, we have the following asymptotic relation $\log(1-x)=-x+O(x^2)$. We thus have
\begin{align}\label{eq:eq 2}
   & K\log\left(  1 - \exp{\left( -\eta^{\frac{\alpha}{M-N_R}-1}\log\eta +\frac{1}{\eta} \right)}  \left( \Theta\left(\frac{1}{\eta^{\alpha}\left( \log\eta \right)^{M-N_R} } \right) + O\left( \frac{1}{\left( \eta^{\frac{\alpha}{M-N_R}} \log\eta \right)^{M-N_R+1}} \right) \right) \right) \notag \\
   & = - \Theta\left(\frac{K}{\eta^{\alpha}(\log\eta)^{M-N_R}}\right) \exp{ \left(-\eta^{\frac{\alpha}{M-N_R}-1}\log\eta +\frac{1}{\eta} \right)} + O\left(\frac{K}{\left( \eta^{\frac{\alpha}{M-N_R}} \log\eta \right)^{M-N_R+1}}\right) \exp{ \left(-\eta^{\frac{\alpha}{M-N_R}-1}\log\eta +\frac{1}{\eta} \right)} \notag \\
   &  + O\left( \Theta\left(\frac{K}{\eta^{2\alpha}(\log\eta)^{2n}}\right) \exp{ \left(-2\eta^{\frac{\alpha}{n}-1}\log\eta +\frac{2}{\eta} \right)} + O\left(\frac{K}{\left( \eta^{\frac{\alpha}{n}} \log\eta \right)^{2(n+1)}}\right) \exp{ \left(-2\eta^{\frac{\alpha}{n}-1}\log\eta +\frac{2}{\eta} \right)} \right) \notag \\
   & \xrightarrow{\eta\to\infty}0,
\end{align}
since $K=\Theta(\eta^{\alpha})$, and $0<\alpha\leq M-N_R$. As a consequence, the upper-bound probability converges to 1 when $\eta\to\infty$. To prove the convergence to 1 of the
lower-bound probability in (\ref{eq:eq 1}), we observe that
\begin{align}\label{eq:eq 3}
    & \text{Pr}\left\{\frac{\alpha}{M-N_R}\log_2\eta - \log_2\log\eta\geq \max_{k \in \{ 1, \cdots, K\}}Y_k \right\}
    = \bigg[ F_{Y_k}\left( \frac{\alpha}{M-N_R}\log_2\eta - \log_2\log\eta \right) \bigg]^K \notag \\
    & = \Bigg(  1 - \exp{\left( -\eta^{\frac{\alpha}{M-N_R}-1}\frac{1}{\log\eta} +\frac{1}{\eta} \right)} \left( \Theta\left(\frac{\left( \log\eta \right)^{M-N_R}}{\eta^{\alpha}} \right) + O\left( \frac{\left( \log\eta \right)^{M-N_R+1}}{ \eta^{\frac{\alpha(M-N_R+1)}{M-N_R}}} \right) \right) \Bigg)^K.
\end{align}
Note that
\begin{align}\label{eq:eq 4}
    & K\log\left(  1 - \exp{\left( -\eta^{\frac{\alpha}{M-N_R}-1}\frac{1}{\log\eta} +\frac{1}{\eta} \right)} \left( \Theta\left(\frac{\left( \log\eta \right)^{M-N_R}}{\eta^{\alpha}} \right) + O\left( \frac{\left( \log\eta \right)^{M-N_R+1}}{ \eta^{\frac{\alpha(M-N_R+1)}{M-N_R}}} \right) \right) \right)    \notag \\
    & = - \Theta\left(\frac{K(\log\eta)^{M-N_R}}{\eta^{\alpha}}\right) \exp{ \left(-\eta^{\frac{\alpha}{M-N_R}-1}\frac{1}{\log\eta} +\frac{1}{\eta} \right)} + O\left(\frac{K\left( \log\eta \right)^{M-N_R+1}}{\eta^{\frac{\alpha(M-N_R+1)}{M-N_R}}}\right) \exp{ \left(-\eta^{\frac{\alpha}{M-N_R}-1}\frac{1}{\log\eta} +\frac{1}{\eta} \right)} \notag \\
    &  + O\left( \Theta\left(\frac{K(\log\eta)^{2(M-N_R)}}{\eta^{2\alpha}}\right) \exp{ \left(-2\eta^{\frac{\alpha}{M-N_R}-1}\frac{1}{\log\eta} +\frac{2}{\eta} \right)} + O\left(\frac{K\left( \log\eta \right)^{2(M-N_R+1)}}{\eta^{\frac{2\alpha(M-N_R+1)}{M-N_R}}}\right) \exp{ \left(-2\eta^{\frac{\alpha}{M-N_R}-1}\frac{1}{\log\eta} +\frac{2}{\eta} \right)} \right) \notag \\
    & \xrightarrow{\eta\to\infty}-\infty,
\end{align}
since, when $\eta\to\infty$, the first term goes to $-\infty$, while the second term goes to 0. (\ref{eq:eq 3}) thus converges to 0 and the lower-bound probability is confirmed. We omit the proof of (\ref{eq:proof01_1b}) since it follows similar arguments. With (\ref{eq:proof01_1a}) and (\ref{eq:proof01_1b}), the results in (\ref{case:theoDoF single-cell RBF-MMSE a}) and (\ref{case:theoDoF single-cell RBF-MMSE b}) follow immediately. 

\subsubsection{Case 2, $N_R \geq M$}\label{subproof:theoDoF single-cell case 2}
Suppose that $M$ receive antennas are used. Then the DoF is $M$ from Case 1 above. Therefore, $d_{\text{RBF-MMSE}}(\alpha,\mv{m}) \geq M$. Also note that in a single-cell MIMO RBF with $M$ transmit beams, the BS can be considered as having $M$ transmit antennas only. Proposition \ref{prop:single-cell DoF upperbound} thus leads to $d_{\text{RBF-MMSE}}(\alpha,\mv{m}) \leq M$. We thus conclude that $d_{\text{RBF-MMSE}}(\alpha,\mv{m}) = M$. 

\subsection{RBF-MF/AS}
To obtain the DoF of RBF-MF/AS, we first show that
\begin{align}\label{eq:proof_2a}
    \text{Pr}\left\{\frac{\alpha}{M-1}\log_2\eta +\log_2\log\eta\geq
    \max_{k \in \{ 1, \cdots, K\}}R_{k,1}^{(\text{MF/AS})} \right.
    \left. \geq \frac{\alpha}{M-1}\log_2\eta -\log_2\log\eta \right\} 
    \xrightarrow{\eta\to\infty}1,~ \text{if}~ 0< \alpha\leq M-1,
\\ \label{eq:proof_2b}
    \text{Pr}\bigg\{ \log_2\eta +\log_2\log\eta +\log_2\alpha\geq
    \max_{k \in \{ 1, \cdots, K\}}R_{k,1}^{(\text{MF/AS})}
    \geq\log_2\eta +\log_2\log\eta +\log_2\beta \bigg\} 
    \xrightarrow{\eta\to\infty}1,~ \text{if}~ \alpha> M-1,
\end{align}
where $\beta_2=\frac{\alpha-M+1}{2}$; hence, $\alpha>\beta_2>0$ when $\alpha>M-1$. From Theorem \ref{theorem:SINR RBF-MF CDF}, the CDF of the single-cell RBF-MF $S:=\text{SINR}_{k,m}^{(\text{MF})}$ is
\begin{align}
F_S(s) = 1 - e^{-s/\eta} \displaystyle\sum_{k=0}^{N_R-1} \sum_{m=0}^k \frac{s^k} { (k-m)! m!  \eta^{k-m}} \frac{\frac{(M+m-2)!}{(M-2)!}}{(s+1)^{M+m-1}}.
\end{align} 
Denote $Z_k$ $:=$ $R_{k,1}^{(\text{MF})} := \log_2\left(1+\text{SINR}_{k,1}^{(\text{MF})}\right)$. The CDF of $Z_k$ is thus 
\begin{align}
F_{Z_k}(z) = 1 - e^{-z/\eta} \displaystyle\sum_{k=0}^{N_R-1} \sum_{m=0}^k \frac{(M+m-2)!} { (k-m)! m! (M-2)!}  \frac{(2^z-1)^k}{\eta^{k-m}2^{(M+m-1)z}}.
\end{align}
In (\ref{eq:proof_2a}), the upper-bound 
probability can thus be given as
\begin{align}\label{eq:eq 5}
    & \text{Pr}\left\{\frac{\alpha}{M-1}\log_2\eta +\log_2\log\eta\geq \max_{k \in \{ 1, \cdots, K\}}Z_k \right\}
    = \bigg[ F_{Z_k}\left( \frac{\alpha}{M-1}\log_2\eta +\log_2\log\eta \right) \bigg]^K \notag \\
    & = \Bigg(  1 - \exp{\left( -\eta^{\frac{\alpha}{M-1}-1}\log\eta +\frac{1}{\eta} \right)}  \left( \displaystyle\sum_{k=0}^{N_R-1} \sum_{m=0}^k \frac{(M+m-2)!} { (k-m)! m! (M-2)!}  \frac{(\eta^{\frac{\alpha}{M-1}}\log\eta-1)^k}{\eta^{k-m}\eta^{\frac{\alpha(M+m-1)}{M-1}}(\log\eta)^{(M+m-1)}} \right) \Bigg)^K \notag \\
    & = \Bigg(  1 - \exp{\left( -\eta^{\frac{\alpha}{M-1}-1}\log\eta +\frac{1}{\eta} \right)}  \left( \Theta\left(\frac{1}{\eta^{\alpha}\left( \log\eta \right)^{M-1} } \right) + O\left( \frac{1}{ \eta^{\frac{\alpha M}{M-1}} } \right) \right) \Bigg)^K,
\end{align}
as $\eta\to\infty$, which is quite similar to (\ref{eq:eq 1}) with $N_R$ = 1 in this case. Now following the same reasoning as in (\ref{eq:eq 2}), we can prove that the upper-bound probability (\ref{eq:eq 5}) $\to 1$ as $\eta\to\infty$. To prove the convergence of the lower-bound, we note that
\begin{align}\label{eq:eq 6}
    & \text{Pr}\left\{\frac{\alpha}{M-1}\log_2\eta - \log_2\log\eta\geq \max_{k \in \{ 1, \cdots, K\}}Z_k \right\}
    = \bigg[ F_{Z_k}\left( \frac{\alpha}{M-1}\log_2\eta - \log_2\log\eta \right) \bigg]^K \notag \\
    & = \Bigg(  1 - \exp{\left( -\eta^{\frac{\alpha}{M-1}-1}\frac{1}{\log\eta} +\frac{1}{\eta} \right)}  \left( \displaystyle\sum_{k=0}^{N_R-1} \sum_{m=0}^k \frac{(M+m-2)!} { (k-m)! m! (M-2)!}  \frac{(\eta^{\frac{\alpha}{M-1}}\frac{1}{\log\eta}-1)^k (\log\eta)^{M+m-1}} {\eta^{k-m}\eta^{\frac{\alpha(M+m-1)}{M-1}}} \right) \Bigg)^K \notag \\
    & = \Bigg(  1 - \exp{\left( -\eta^{\frac{\alpha}{M-1}-1}\frac{1}{\log\eta} +\frac{1}{\eta} \right)}  \left( \Theta\left(\frac{\left( \log\eta \right)^{M-1}}{\eta^{\alpha} } \right) + O\left( \frac{(\log\eta)^{M-1}}{ \eta^{\frac{\alpha M}{M-1}} } \right) \right) \Bigg)^K,
\end{align}
which is quite similar to (\ref{eq:eq 3}). Now following the same reasoning as in (\ref{eq:eq 4}), we can prove that (\ref{eq:eq 6}) $\to 0$ as $\eta\to\infty$. Thus we confirm (\ref{eq:proof_2a}). The proof of (\ref{eq:proof_2b}) follows similarly and is thus omitted. 

On the other hand, for the case of RBF-AS, note that RBF-AS scheme consists of two selection processes: antenna selection at each MS with $N_R$ antennas and user selection at the BS with $K$ users. The rate performance of RBF-AS is therefore equivalent to that of MISO RBF with $N_RK$ single-antenna users in the cell. Thus, we obtain (\ref{eq:proof_2a}) and (\ref{eq:proof_2b}) for the case of RBF-AS. With (\ref{eq:proof_2a}) and (\ref{eq:proof_2b}), the results in (\ref{case:theoDoF single-cell RBF-MF/AS a}) and (\ref{case:theoDoF single-cell RBF-MF/AS b}) follow immediately. 

This thus completes the proof of Lemma \ref{lemma:theoDoF single-cell RBF-MMSE, RBF-MF/AS}.

\section{Proof of Proposition \ref{prop:single-cell DoF upperbound}}\label{proof:single-cell DoF upperbound}
In a single-cell MIMO-BC, DPC yields the maximum sum-rate, denoted by $R_{\text{DPC}}$. Therefore, the single-cell DoF can be bounded as $d \leq$ $\displaystyle\lim_{\rho\to\infty}\frac{R_{\text{DPC}}}{\log_2\rho}$. From \cite[Theorem 1]{Jindal01}, we have
\begin{align}\label{eq:single-cell DoF bounded}
R_{\text{DPC}} 
& \leq N_T\mathbb{E}\left[ \log_2\left[ 1 + \eta\max_{k \in \{ 1, \cdots, K\}}||\mv{H}_{k}||_2^2 \right] \right] \notag \\
& \leq N_T\mathbb{E}\left[ \log_2\left[ 1 + \eta\max_{k \in \{ 1, \cdots, K\}}\mv{Tr}\left(\mv{H}_{k}^H\mv{H}_{k}\right) \right] \right].
\end{align}
Denote $R_k$ $:=$ $\log_2\left[ 1 + \eta\mv{Tr}\left(\mv{H}_{k}^H\mv{H}_{k}\right) \right]$. Note that $\mv{Tr}\left(\mv{H}_{k}^H\mv{H}_{k}\right)$ is distributed as $\chi^2(2N_TN_R)$. Similarly to (\ref{eq:proof01_1b}) and (\ref{eq:proof_2b}), we can show that
\begin{align}\label{eq:bound 2}
	\text{Pr}\left\{ \log_2\eta + \log_2\log\eta + \log_2(\alpha+1)\geq \max_{k \in \{ 1, \cdots, K\}}R_k\right\} \xrightarrow{\eta\to\infty}1.
\end{align}
Combining (\ref{eq:single-cell DoF bounded}) and (\ref{eq:bound 2}), we obtain $d\leq N_T$, where the equality is achieved by, e.g., the DPC scheme. The proof of Proposition \ref{prop:single-cell DoF upperbound} is thus completed. 



\end{document}